\documentclass[preprintnumbers,prd,showpacs,floatfix,superscriptaddress,nofootinbib,twocolumn, letterpaper]{revtex4-1}
\pdfoutput=1

\usepackage{longtable}
\usepackage{bm}
\usepackage{relsize}
\usepackage{amsfonts}
\usepackage{amsmath}
\usepackage{amssymb,epsf}
\usepackage{latexsym}
\usepackage{graphicx,epsfig}
\usepackage{amssymb}
\usepackage{float}
\usepackage{subfigure}
\usepackage{epstopdf}
\usepackage[colorlinks=true,citecolor=blue,linkcolor=blue,urlcolor=black]{hyperref}
\usepackage{dcolumn}
\usepackage{psfrag}
\usepackage{wrapfig}
\usepackage{makeidx}
\usepackage{epsf}
\usepackage{color}
\usepackage{multirow}
\usepackage{mathtools}

\def\InstOfThPhyAstroUG{Institute of Theoretical Physics and Astrophysics,
University of Gdańsk,
80-308 Gda\'{n}sk, Poland}

\def\Tartu{Institute of Physics, University of Tartu, W. Ostwaldi 1, 50411 Tartu, Estonia}

\begin{document}

\title{Dynamical system analysis of cosmological evolution in the Aether scalar tensor theory}

\author{João Luís Rosa}
\email{joaoluis92@gmail.com}
\affiliation{\Tartu}
\affiliation{\InstOfThPhyAstroUG}

\author{Tom Zlosnik}
\email{thomas.zlosnik@ug.edu.pl}
\affiliation{\InstOfThPhyAstroUG}

\date{\today}

\begin{abstract} 

The Aether Scalar Tensor (AeST) theory is an extension of General Relativity (GR), proposed for addressing galactic and cosmological observations without dark matter.
The action for the theory includes a function that can currently only be constrained by phenomenological considerations. In antecedent work, forms of this function were considered that led to an effective fluid contribution to the cosmological evolution equations that approximated that of dust more and more closely at late cosmic times. In this work we consider an alternative set of functions that most closely approximate dust at the earliest cosmic times and where deviations from dust-like behaviour gradually emerge with time. We use the dynamical system formalism to analyze example models from both possible sets of functions, introducing a complete set of dynamical variables describing the spacetime curvature, energy density parameters of different matter components, and AeST scalar field, and obtain the dynamical equations describing cosmological evolution. The cosmological phase space is found to feature invariant submanifolds associated to the absence of the matter components, as well as equilibrium states associated with well-known cosmological behaviors e.g. matter, radiation, and cosmological constant dominated epochs. A full numerical integration of the dynamical system is performed for the models and it is shown that each can closely approximate the $\Lambda \mathrm{CDM}$ model at the level of the cosmic background. Generalizations of the models are considered and it is shown that the new models likely can simultaneously replicate the cosmological successes of cold dark matter whilst satisfying constraints on the theory from the weak-field quasistatic regime. 
\end{abstract}

\maketitle

\section{Introduction}\label{sec:intro}

Despite the emergence of evidence for dark matter across a wide range of astrophysical and cosmological systems, this evidence is entirely via dark matter's effect on visible matter. Furthermore, there is evidence that the total gravitational field in many astrophysical systems in the presence of dark matter is correlated with that that would be due to baryons in the absence of dark matter \cite{McGaughEtAl2016}. This behaviour can be described in terms of a modification to Newton's theory of gravity \cite{Milgrom1983a}:

\begin{align}
\vec{\nabla}\cdot\bigg(\mu\bigg(\frac{|\vec{\nabla}\Phi|}{a_{M}}\bigg)\vec{\nabla}\Phi\bigg) &=  \nabla^{2}\Phi_{B} = 4\pi G\rho_{B} \label{modified_poisson}
\end{align}
Where $\Phi$ is the Newtonian gravitational potential, $\mu(x)$ is a function which is not fixed by the theory beyond having limiting forms $\mu(x)\sim x$ for $x\ll 1$ and $\mu(x) \sim 1$ for $x \gg1$, $a_{M} \sim 10^{-10} m s^{-2}$ is an empirically-determined scale with dimensions of acceleration, and $\Phi_{B}$ is the gravitational potential that would correspond to a baryonic density distribution $\rho_{B}$. It is conceivable that Eq. (\ref{modified_poisson}), rather than reflecting an occasional strong correlation between baryonic and dark matter influence on the gravitational field, instead represents a limit of a larger purely gravitational theory (containing no dark matter) much in the same way that Newtonian gravity arises as limiting behaviour in General Relativity. This is the paradigm of Modified Newtonian Dynamics (MOND) and a number of variations on Eq.
(\ref{modified_poisson}) additionally exist (for example involving additional gravitational potentials), which nonetheless can lead to comparable phenomenology (see for example \cite{MOND-QLinear-Milgrom2009,Zhao:2012ky,Milgrom:2023idw}).  MOND has been confronted with an extensive amount of astrophysical data \cite{MOND-Famaey2012,Hees:2015bna}.

A number of proposals \cite{BekensteinMilgrom1984,Bekenstein1988,Sanders1997,Bekenstein2004,NavarroVanAcoleyen2005,ZlosnikFerreiraStarkman2006,Sanders2007,Milgrom2009,BabichevDeffayetEsposito-Farese2011,DeffayetEsposito-FareseWoodard2011,Mendoza:2012hu,Woodard2014,Khoury2014,Blanchet:2015sra,Hossenfelder2017,Burrage:2018zuj,Milgrom:2019rtd,DAmbrosio:2020nev,Kading:2023hdb} have been put forward for extensions to General Relativity that recover Eq.(\ref{modified_poisson}) - or variations thereof - in appropriate limits. One recent proposal - the AeST (Aether Scalar Tensor) model - recovers the correct modification to Newton's theory in the quasistatic, weak-field limit whilst providing a similarly good match to data from the cosmic microwave background radiation and large scale structure as in the standard $\Lambda\mathrm{CDM}$ ($\Lambda$ cold dark matter) cosmological model \cite{SkordisZlosnik2020}. Various properties of the AeST model have been investigated in further work ~\cite{Bernardo:2022acn,Kashfi:2022dyb,Mistele:2021qvz,Mistele:2023paq,Tian:2023gjt,Llinares:2023lky,Verwayen:2023sds,Mistele:2023fwd,Bataki:2023uuy}.

Part of this success of the AeST model may be attributed to one of its degrees of freedom providing an additional dust-like contribution to the cosmological background stress energy tensor, akin to the dark matter paradigm. However, an interesting feature of this model is that the dust-like contribution generally occurs for a limited period of cosmic time. Analogous to the functional freedom present in the function $\mu(x)$ in Eq. (\ref{modified_poisson}), AeST contains a function ${\cal F}$ whose exact form is not determined by basic principles. Different forms of this function can lead to different phenomenology. In \cite{SkordisZlosnik2020}, several functions were considered that lead to an additional effective contribution to the cosmological matter density which scaled as dust at late cosmic times but at earlier times could scale alternatively, for example as an additional radiation component.

\begin{figure}
    \includegraphics[height=5cm]{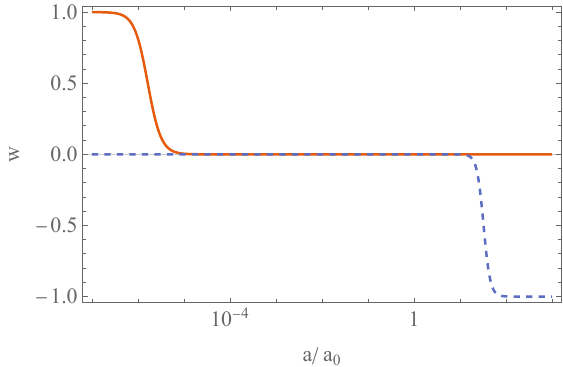}
    \caption{The evolution of the equation of state $w$ of the effective fluid contribution to the Einstein equations due to AeST's new gravitational degrees of freedom as a function of the ratio of the scale factor $a$ to its present-day value $a_{0}$ is shown for examples of Model 1 (solid line, see Section \ref{classA}) and Model 2 (dashed line, see Section \ref{classB}). Both share an extended span of time where $w\sim 0$ but differ significantly at early and late times.}
    \label{fig:wofa}
\end{figure}

In this paper we will look at a wider class of functions ${\cal F}$, in particular introducing functions which mostly closely approximate dust at earlier cosmic time and deviate from dust-like behaviour at \emph{late} cosmic times (with signifant deviations potentially arising only in the far cosmic future). The equation of state of the effective fluid associated with ${\cal F}$ for two such functions is shown in Figure \ref{fig:wofa}. For a set of relevant functions, we will conduct a dynamical system analysis of the equations describing background cosmic evolution, characterize how the expansion history compares to $\Lambda\mathrm{CDM}$, examine the implications for the evolution of cosmic perturbations and the effect on phenomenology in the (modified)-Newtonian regime, which is sensitive to the form ${\cal F}$ takes cosmologically. The methods we employ have been previously applied to several extensions of GR \cite{Odintsov:2017tbc,Carloni:2015jla,Alho:2016gzi,Carloni:2007eu,Carloni:2017ucm,Carloni:2009jc,Carloni:2015lsa,Carloni:2015bua,Tamanini:2013ltp,Rosa:2019ejh,Carloni:2018yoz,Carloni:2007br,Carloni:2013hna,Bonanno:2011yx,Goncalves:2023klv}, thus emphasizing their versatility and richness. We also refer to \cite{Wiggins2003,Perko2001} for extensive textbooks on the topic, and also to \cite{Bahamonde:2017ize} for a review in dynamical systems applied to cosmology.

The outline of the paper is as follows: In Section \ref{sec:theory} we briefly survey the AeST model and its equations of motion in Friedmann-Lema\^{i}tre-Robertson-Walker (FLRW) symmetry, with an emphasis on how AeST's fields can give rise to an effective dust-like component. In Section \ref{sec:dynsys} we present the framework of dynamical systems analysis and in Section \ref{sec:fixed_points} we determine the nature of fixed points in the dynamical system phase space for a number of models of interest. In Section \ref{sec:cosmo_models} we determine whether 
these models can produce a cosmological expansion history similar to that in the standard cosmological model. In Section \ref{sec:generalizations} we generalize the models analyzed in the prior sections and determine what constraints exist upon models so that they produce a viable alternative to dark matter at the level of cosmological perturbations whilst satisfying astrophysical constraints on the theory. In Section \ref{sec:concl} we present our conclusions.

\section{Theory}\label{sec:theory}
%
The action for the AeST model is as follows:

\begin{widetext}
\begin{align}
    S &=  \int d^{4}x\sqrt{-g}\frac{1}{16\pi \tilde{G}}\bigg(R-\frac{K_{B}}{2}F_{ab}F^{ab}+2(2-K_{B})J^{a}\nabla_{a}\phi -(2-K_{B})Y-{\cal F}(Y,Q)- \lambda(g_{ab}A^{a}A^{b}+1)\bigg)+S_{m}[g] \label{aest_action}
\end{align}
\end{widetext}
The action is a functional of fields $(g_{ab},A_{a},\phi,\lambda)$  where $\lambda$ acts as a Lagrange multiplier field to implement a fixed-norm constraint on $A_{a}$, where $R$ is the Ricci scalar corresponding to $g_{ab}$, and

\begin{align}
 F_{ab} &\equiv 2\partial_{[a}A_{b]}\\
 J^{a} &\equiv A^{b}\nabla_{b}A^{a}\\
Y &\equiv (g^{ab}+A^{a}A^{b})\partial_{a}\phi \partial_{b}\phi \\
 Q &\equiv A^{a}\partial_{a}\phi
 \end{align}
and $K_{B}$ is a dimensionless constant. We will be interested in the behaviour of the theory in spacetimes with FLRW symmetry, in which case $g_{ab}dx^{a}dx^{b} = -N^{2}dt + a^{2}(t)h_{ab}dx^{a}dx^{b}$, $A^{a}\partial_{a} = N^{-1}\partial_{t}$ and $\phi=\phi(t)$, where $h_{ab}$ is a metric of a maximally symmetric 3-space. Note that this implies that here $Y=0$ and $Q=\dot{\phi}/N$, where a dot denotes a derivative with respect to the time coordinate. The correct equations of motion can be recovered from an action adapted to this symmetry \cite{Fels:2001rv}:
\begin{align}
S &=  \frac{1}{8\pi \tilde{G}} \int dtd^{3}x N a^{3}\bigg(-\frac{3H^{2}}{N^{2}}+F(Q)\bigg) + S_{m}[g] \label{aest}
\end{align}
where we have defined $F(Q)=-\frac{1}{2}{\cal F}(0,Q)$ and $H=\dot{a}/a$. Note that the constant $\tilde{G}$ may differ in small proportion from the measured value of Newton's constant but the difference may be sufficiently small so that the values can be taken to essentially coincide \cite{SkordisZlosnik2020}. Adopting the proper time gauge $N=1$, the Einstein equations then take the form:

\begin{equation}
H^2+\frac{k}{a^2}=\frac{8\pi\tilde{G}\rho}{3}-\frac{1}{3}\left(F-QF_Q\right)+\frac{\Lambda}{3}, \label{ein1}
\end{equation}
\begin{equation}
2\dot H+3H^2+\frac{k}{a^2}=-8\pi\tilde{G} p-F+\Lambda, \label{ein2}
\end{equation}
where $F_{Q}\equiv dF/dQ$ and we have allowed for matter content described by collective density $\rho$ and pressure $p$ which may be composed of a number of contributions (e.g. baryon and radiation density); $\Lambda$ corresponds to a cosmological constant. The equation of motion for the scalar field takes the form
\begin{equation}\label{eq:eomQ}
\frac{dF_{Q}}{dt}+3HF_Q=0.
\end{equation}
Each matter component described by density $\rho_{(i)}$ and pressure $p_{(i)}$ obeys the standard conservation equation:
\begin{equation}\label{eq:conservation}
\dot\rho_{(i)}+3H\left(\rho_{(i)}+p_{(i)}\right)=0.
\end{equation}
To model the cosmological background of our universe, we consider that the matter distribution consists of two perfect fluids, i.e., $\rho=\rho_m+\rho_r$ and $p=p_m+p_r$. The first fluid represents a pressureless matter component (dust), satisfying the equation of state $p_m = \omega_m \rho_m$, with $\omega_m=0$, and the subscript $m$ stands for matter. The second fluid represents a radiation component satisfying the equation of state $p_r=\omega_r\rho_r$, with $\omega_r=1/3$, and the subscript $r$ stands for radiation. Furthermore, we assume that both fluids are independently conserved, i.e., these two fluids thus satisfy the following conservation equations
\begin{equation}\label{eq:cons_m}
\dot \rho_m+3H \rho_m=0,
\end{equation}
\begin{equation}\label{eq:cons_r}
\dot \rho_r+4H\rho_r=0,
\end{equation}
Finally we note that the degree of freedom $\phi$ can be cast as a perfect fluid; its appearance in the Einstein equations, Eqs. (\ref{ein1}) and (\ref{ein2}) suggests an identification
\begin{align}
    \rho_{(\phi)} &\equiv -\frac{1}{8\pi \tilde{G}}(F-QF_{Q}), \\
    P_{(\phi)} &\equiv \frac{1}{8\pi \tilde{G}}F,
\end{align}
and indeed it can be shown that for non-zero $Q$, the scalar field equation of motion Eq.(\ref{eq:eomQ}) is equivalent to the fluid conservation equation Eq. (\ref{eq:conservation}). We can further introduce the equation of state of the field $\phi$ and its adiabatic sound speed $c_{ad(\phi)}^{2}$ which will prove to be useful:
\begin{align}
    w_{(\phi)} &=  \frac{P_{(\phi)}}{\rho_{(\phi)}} = \frac{F}{-F+QF_{Q}},  \\
    c_{ad(\phi)}^{2} &= \frac{dP_{(\phi)}}{d\rho_{(\phi)}} = \frac{F_{Q}}{QF_{QQ}}.
\end{align}
The effective fluid associated with the field $\phi$ produces a gravitational effect similar to that of dust if
\begin{align}
\rho_{(\phi)} &>0   \quad  \big(QF_{Q}-F>0\big),\\
|w_{(\phi)}| &\ll 1  \quad \big(|F| \ll QF_{Q}-F\big). \label{wphi_con}
\end{align}
Now we consider $Q$ to be close to a fixed value $Q_{0}$ so that we can perform a series expansion:
\begin{align}
F(Q) &= (Q-Q_{0})^{n} + \dots \label{f_exp}.
\end{align}
Given the condition in Eq. (\ref{wphi_con}), we should have $n>0$ and $\dots$ denote terms of higher order in $(Q-Q_{0})$. Using Eq. (\ref{f_exp}) we have that:
\begin{align}
w_{(\phi)} &= \frac{Q-Q_{0}}{nQ - (Q-Q_{0})} + \dots,\\
\rho_{(\phi)} &= 8\pi \tilde{G} (Q-Q_{0})^{n-1}(nQ - (Q-Q_{0})) + \dots.
\end{align}
Then we have - making use of the integrated equation of motion Eq. (\ref{eq:eomQ}):
\begin{align}
n(Q-Q_{0})^{n-1} &= \frac{\xi}{a^{3}} + \dots \label{inv_n},
\end{align}
where $\xi$ is a constant of integration. Hence we have 
\begin{align}
\rho_{(\phi)} &= \frac{(8\pi \tilde{G}n\xi Q_{0})}{a^{3}}+ \dots
\end{align}
Therefore for $Q$ close to $Q_{0}$, $\rho_{(\phi)}$ gravitates - up to small corrections - as a dust-like component if $n\neq 1$ (from Eq. (\ref{inv_n})) and $(c,Q_{0})\neq 0$. We see from Eq. (\ref{inv_n}) that the value of $n$ determines whether this dark matter type effect is approached as $a$ becomes smaller or larger. In \cite{SkordisZlosnik2020} a number of examples were considered with $n=2$ which have $Q$ approaching $Q_{0}$ for large $a$. In this paper we consider examples where $n=1/2$, and therefore the smaller the value of $a$ is, the closer $w_{(\phi)}$ is to zero. Though the presence of non-analytic functions of gradients of fields in a model's action may appear unusual, they are common in field theories of MOND \cite{BekensteinMilgrom1984}.
In Section \ref{sec:generalizations} we show that this difference can have important consequences for constraining the theory.

Might a function $F(Q)$, which may be approximated by the expansion in Eq. (\ref{f_exp}), have a deeper theoretical origin? The case $n=2$ matches leading order derivative terms in FRW symmetry in the case of the ghost condensate scalar field model \cite{Arkani-Hamed:2003pdi} which itself, for example, might arise as a low-energy effective action arising from a renormalizable theory of a complex scalar field coupled to a large number of massive fermions \cite{Graesser:2005ar}. The case $n=1/2$ matches the leading order derivative terms in FRW symmetry for the Born-Infeld scalar field model, which arises as a limiting form of effective actions describing a particular scalar degree of freedom in the context of string theory \cite{Sen:2002an}. 

The AeST model represents a generalization of models of gravity coupled to a scalar field with non-canonical kinetic term (commonly referred to as \emph{K-essence} models \cite{Armendariz-Picon:1999hyi}), and this is due to the presence of the field $A^{a}$. Generally it is possible to define a new field $B^{a}=\phi A^{a}$ (so that the degree of freedom $\phi$ measures the norm of $B^{a}$) and write the action in Eq. (\ref{aest_action})  as a model of gravity coupled to a vector field with non-canonical kinetic term. By comparison we then have:

\begin{align}
S[g,\phi] &= \frac{1}{16\pi \tilde{G}} \int \sqrt{-g}\bigg(R + K(\phi,\nabla\phi)\bigg) \quad (\mathrm{K-essence}) \label{k_action}\\
S[g,B^{a}] &= \frac{1}{16\pi \tilde{G}}\int \sqrt{-g}\bigg(R+  K(B,\nabla B)\bigg) \quad  (\mathrm{AeST)} \label{gB_action}
\end{align}
The ghost-condensate and Born-Infeld models represent examples of Eq. (\ref{k_action}) which are limiting forms of theories with a deeper theoretical underpinning. The class of actions in Eq. (\ref{gB_action}) - which includes the AeST model in Eq. (\ref{aest_action}) in the sense of producing the same equations of motion - may be considered as a generalization of Eq. (\ref{k_action}) to the case where the action depends on more components of the vector field than simply the norm $\phi^{2}=-g_{ab}B^{a}B^{b}$. Whether such generalizations of Eq. (\ref{k_action}) can be shown to arise from more fundamental theory remains an open problem.

\section{Dynamical system analysis}\label{sec:dynsys}

\subsection{Brief review of dynamical systems}

The details concerning the methods of the dynamical system approach can be found in several mathematics textbooks \cite{Wiggins2003,Perko2001}. Nevertheless, to preserve the self-consistency of this work, we briefly review the essentials of the methods necessary to follow the forthcoming analysis. Consider a dynamical system of coupled differential equations for a given set of $n$ dynamical variables $X_i(x)$, where $x$ is an independent variable such that each variable satisfies an equation of the form
\begin{equation}
X'_i(x)=f_i\left(X_1(x),...,X_n(x)\right),
\end{equation}
where a prime denotes a derivative with respect to $x$ and $f_i$ are $n$ well-behaved functions of the quantities $X_i$. A point in the phase space at which all derivatives of the dynamical variables vanish, i.e., $X'_i(x)=f_i\left(X_1^{P}(x),...,X_n^{P}(x)\right)=0$, where $X_i^{P}$ denote the values of the dynamic variables at a given point $P$, is dubbed a \textit{fixed point}. These points may present three different types of behaviors: if every trajectory in the phase space passing through a point in the vicinity of $P$ is directed towards $P$, then $P$ is an \textit{attractor}; if every trajectory in the phase space passing through a point in the vicinity of $P$ is directed outwards from $P$ and can be traced backwards to $P$, then $P$ is a \textit{reppeler}; and if none of the conditions above apply then $P$ is a \textit{saddle}. If $P$ is an attractor or a reppeler for every point in the phase space and not just for the points in a finite vicinity of $P$, then $P$ is a \textit{global} attractor or reppeler. 

The behavior of a fixed point can be computed as follows. Consider the following vector field $g$ defined as
\begin{equation}
g\left(X_1,...,X_n\right)=\left(f_1,...,f_n\right).
\end{equation}
The Jacobian matrix $Jg$ of the vector field $g$ can be written as
\begin{equation}
J\left(X_1,...,X_n\right)=\begin{pmatrix}
\frac{\partial f_1}{\partial X_1} & ... & \frac{\partial f_1}{\partial X_n}\\
... & ... & ... \\
\frac{\partial f_n}{\partial X_1} & ... & \frac{\partial f_n}{\partial X_n}
\end{pmatrix}.
\end{equation}
For a given fixed point $P$ with coordinates $X_i^P$, compute $J^P\equiv J\left(X_1^P,...,X_n^P\right)$. The eigenvalues of the matrix $J^P$ are denoted as $\lambda_i$. If the matrix $J^P$ is positively defined, i.e., if $\lambda_i>0$, the fixed point $P$ is a reppeler; If the matrix $J^P$ is negatively defined, i.e., if $\lambda_i<0$, the fixed point $P$ is an attractor; If the matrix $J^P$ is indefinite, i.e., if one has $\lambda_i>0$ and $\lambda_j<0$ for some $\{i,j\}\in\{1,...,n\}$, the fixed point $P$ is a saddle. If the matrix $J^P$ is semi-definite, i.e., if the eigenvalues are $\lambda_i\geq 0$ or $\lambda_i\leq 0$, this analysis is not sufficient to draw conclusions about the behavior of the fixed point, and one needs to recur to alternative methods. One such method is the central manifold analysis \cite{Wiggins2003}, which provides a robust mathematical analysis of the stability of a fixed point. In the analysis that follows, in some cases we find fixed points for which the matrix $J^P$ is semi-definite. However, since the stability of these fixed points can be inferred directly from the analysis of the trajectories on the phase space, and thus the central manifold analysis is not required, we choose to skip these details.

Finally, if for a given variable $X_i$ one has that $X_i=X_i^P$ implies $X_i'=0$, the submanifold $X_i=X_i^P$ is dubbed an \textit{invariant submanifold}, i.e., any trajectory on the phase space that starts from an initial condition that contains $X_i=X_i^P$ is restricted to evolve within that same submanifold. This implies that invariant submanifolds split the phase space into two different regions that can not be accessed by the a single trajectory, as the invariant submanifolds can not be crossed. Due to the presence of invariant submanifolds, a given point $P$ can only be a global attractor or reppeler if it stands in the intersection of all invariant submanifolds.

\subsection{Definitions and equations}

The first step towards a dynamical system analysis of a given system of equations is to write all equations in a dimensionless form. For this purpose, a set of dimensionless dynamical variables representing the relevant quantities describing the system must be defined. We thus introduce the following dynamical variables,
\begin{equation}
K=\frac{k}{a^2H^2},\qquad \Omega_m=\frac{8\pi\rho_m}{3H^2},\qquad \Omega_r=\frac{8\pi\rho_r}{3H^2},\nonumber 
\end{equation}
\begin{equation}\label{eq:dyn_vars}
\Omega_\Lambda=\frac{\Lambda}{3H^2}, \qquad
\Phi=\frac{Q}{H}.
\end{equation}
Furthermore, to allow for the analysis of different particular models for the function $F(Q)$, a set of dimensionless dynamical functions representing this function and its partial derivatives must be defined. These functions are written explicitly in terms of the dynamical variables defined in Eq.\eqref{eq:dyn_vars} upon the specification of a form of $F(Q)$. We thus introduce the following dynamical functions
\begin{equation}\label{eq:dyn_funs}
F_1=-\frac{F}{3H^2},\qquad F_2=-\frac{F_Q}{3H}, \qquad F_3=-\frac{F_{QQ}}{9}.
\end{equation}
Under the definitions above for dimensionless dynamical variables and functions in Eqs.\eqref{eq:dyn_vars} and \eqref{eq:dyn_funs} respectively, the cosmological equations given in Eqs. \eqref{ein1} and \eqref{ein2} take the forms 
\begin{equation}\label{eq:const1}
1+K-\Omega_m-\Omega_r-\Omega_\Lambda-F_1+F_2\Phi=0,
\end{equation}
\begin{equation}\label{eq:const2}
1+K+\Omega_r-3\Omega_\Lambda-3F_1-2q=0,
\end{equation}
where we have introduced the definition of the deceleration parameter $q$, which can be written in terms of the time derivatives of the scale factor and the Hubble parameter as
\begin{equation}\label{eq:defq}
q=-\frac{\ddot a}{a H^2}.
\end{equation}
The two cosmological equations in Eqs.\eqref{eq:const1} and \eqref{eq:const2} represent constraint equations, i.e., they provide algebraic relations between the dynamical variables previously defined, and can be conveniently used to eliminate certain dynamical variables from the dynamical system, thus decreasing the number of dynamical equations necessary to analyze.

To obtain the system of dynamical equations for the variables defined in Eq.\eqref{eq:dyn_vars}, one must introduce a dimensionless time parameter $N$ with respect to which the derivatives of the dynamical variables must be taken. We thus define
\begin{equation}
N=\log\left(\frac{a}{a_0}\right),
\end{equation}
where $a_0$ is an arbitrary constant with units of length. Denoting the derivatives with respect to $N$ by a prime $'$, these derivatives can be transformed into derivatives with respect to the time $t$ via
\begin{equation}\label{eq:Nderivative}
X'=\frac{\dot X}{H}.
\end{equation}

To obtain the dynamical equations describing the dynamical variables defined in Eq.\eqref{eq:dyn_vars}, one takes a derivative with respect to $N$ of each of these variables. Alternatively, for those variables for which the equations of motion feature derivatives with respect to time e.g. Eq.\eqref{eq:eomQ} for $Q$ and the conservation equations in Eqs.\eqref{eq:cons_m} and \eqref{eq:cons_r} for $\rho_m$ and $\rho_r$ respectively, the same dynamical equations can be obtained directly via the transformation of time derivatives into $N-$derivatives with Eq.\eqref{eq:Nderivative}. The dynamical system of equations for the variables defined in Eq.\eqref{eq:dyn_vars} takes the form
\begin{equation}
K'=2qK,
\end{equation}
\begin{equation}
\Omega_m'=\Omega_m\left(2q-1\right),
\end{equation}
\begin{equation}
\Omega_r'=2\Omega_r\left(q-1\right),
\end{equation}
\begin{equation}
\Omega_\Lambda'=2\Omega_\Lambda\left(q+1\right),
\end{equation}
\begin{equation}
\Phi'=\left(1+q\right)\Phi-\frac{F_2}{F_3}.
\end{equation}

The two constraints in Eqs.\eqref{eq:const1} and \eqref{eq:const2} allows one to eliminate two dynamical variables from the system. For convenience, we chose to use Eq.\eqref{eq:const2} to eliminate the deceleration parameter $q$ in terms of the remaining variables. Then we use Eq.\eqref{eq:const1}, which does not depend on $q$, to eliminate the variable $K$. We are thus left with a dynamical system of four equations, namely
\begin{equation}\label{eq:dyn_eq_gen1}
\Phi'=\Phi\left(1-F_1+\frac{1}{2}\Omega_m+\Omega_r-\Omega_\Lambda-\frac{1}{2}F_2\Phi\right)-\frac{F_2}{F_3},
\end{equation}
\begin{equation}\label{eq:dyn_eq_gen2}
\Omega_m'=\Omega_m\left(\Omega_m-1-2F_1-F_2\Phi+2\Omega_r-2\Omega_\Lambda\right),
\end{equation}
\begin{equation}\label{eq:dyn_eq_gen3}
\Omega_r'=\Omega_r\left(\Omega_m+2\Omega_r-2\Omega_\Lambda-2-2F_1-F_2\Phi\right),
\end{equation}
\begin{equation}\label{eq:dyn_eq_gen4}
\Omega_\Lambda'=\Omega_\Lambda\left(\Omega_m+2\Omega_r-2\Omega_\Lambda+2-2F_1-F_2\Phi\right).
\end{equation}
Upon solving the system of Eqs. \eqref{eq:dyn_eq_gen1} to \eqref{eq:dyn_eq_gen4}, the curvature $K$ can be computed from Eq.\eqref{eq:const1}, and afterwards the deceleration parameter $q$ can be computed from Eq.\eqref{eq:const2}. Furthermore, one can identify the presence of three invariant submanifolds in the system, namely $\Omega_m=0$, $\Omega_r=0$, and $\Omega_\Lambda=0$, resulting in $\Omega_m'=0$, $\Omega_r'=0$, and $\Omega_\Lambda'=0$, respectively. This implies that any trajectory in the phase space starting from a set of initial conditions that intersects with one of these invariant submanifolds is restricted to evolve within it. This also implies that any global property of the phase space must lie in the intersection of all three invariant submanifolds, as these submanifolds can not be crossed by trajectories in the phase space. 

To proceed further with the analysis of the structure of the cosmological phase space, it is necessary to define an explicit form of the function $F(Q)$. In what follows, we introduce three well-known classes of models and analyze the fixed points and trajectories of the phase space.

\section{Fixed points and phase space trajectories}
\label{sec:fixed_points}

\subsection{Class 1 model}
\label{classA}

Consider as a first example the following model for the function $F(Q)$,
\begin{equation}\label{eq:model1}
F(Q)=k_0\left(Q-Q_0\right)^2,
\end{equation}
where $k_0$ and $Q_0$ are constant free parameters. While $k_0$ is a dimensionless parameter, and thus no extra dynamical variable is necessary to describe it, the same is not true for the parameter $Q_0$, as the latter features the same units as $Q$. It is thus necessary to define one extra dynamical variable to describe this parameter, namely
\begin{equation}
\Phi_0=\frac{Q_0}{H},
\end{equation}
which satisfies the dynamical equation (obtained by taking a derivative with respect to $N$) 
\begin{equation}\label{eq:dyn_extra1}
\Phi'_0=\Phi_0\left(1+q\right).
\end{equation}
Since $\Phi_0=0$  implies $\Phi'_0=0$, this dynamical equation contributes with an additional invariant submanifold in the cosmological phase space. For this choice of model, the three dynamical functions $F_i$ take the explicit forms
\begin{equation}\label{eq:model1_f1}
F_1=-\frac{1}{3}k_0\left(\Phi-\Phi_0\right)^2,
\end{equation}
\begin{equation}\label{eq:model1_f2}
F_2=-\frac{2}{3}k_0\left(\Phi-\Phi_0\right),
\end{equation}
\begin{equation}\label{eq:model1_f3}
F_3=-\frac{2}{9}k_0.
\end{equation}
It is important to note that, upon replacing Eqs.\eqref{eq:model1_f1} to \eqref{eq:model1_f3} into Eq.\eqref{eq:dyn_eq_gen1}, one verifies that under the assumption $\Phi_0=0$, the submanifold $\Phi=0$ becomes an invariant submanifold, as it implies $\Phi'=0$.

For this model, the system of Eqs.\eqref{eq:dyn_eq_gen1} to \eqref{eq:dyn_eq_gen4}, along with Eq.\eqref{eq:dyn_extra1}, features at most five different isolated fixed points, and a continuous line of fixed points, summarized in Table \ref{tab:fixed1}. One can identify several know behaviors of the scale factor associated with different fixed points, namely the cosmological constant dominated exponential acceleration for points $\mathcal A$, the radiation dominated power-law $\sqrt{t}$ for point $\mathcal B$, the matter dominated power-law $t^\frac{2}{3}$ for point $\mathcal C$, and the scalar field dominated power-law $t^\frac{1}{3}$ for points $\mathcal E^\pm$. These fixed points correspond to spatially-flat spacetimes with $K=0$ and all but $\mathcal E^\pm$ stand in the limit $Q=Q_0$. Note however that $\mathcal A$ exists for any value of $\Phi_0$, whereas points $\mathcal B$, $\mathcal C$ and $\mathcal E^\pm$ only exist for $\Phi_0=0$. Since for $q>-1$ the Hubble parameter $H$ is always finite and vanishes asymptotically, the only possible way for these points to be reached is to consider the particular case $Q_0=0$, in which case the point is reached when $Q=Q_0$. Finally, the point $\mathcal D$ represents a linearly expanding vacuum universe with a closed geometry $K=-1$. Since point $\mathcal D$ stands on the intersection of all invariant submanifolds of the phase space, it is the only fixed point that represents a global property of the phase space. It is also interesting to note that the fixed points $\mathcal E^\pm$ only exist for $k_0\geq0$ and stand at $\Phi=\pm \infty$ for the particular case $k_0=0$, case in which one recovers GR.  

\begin{table}
    \centering
    \begin{tabular}{c|c|c|c|c|c|c|c}
         & $K$ & $\Omega_m$ & $\Omega_r$ & $\Omega_\Lambda$ & $\Phi$ & $\Phi_0$ & $q$  \\ \hline 
         $\mathcal A$ & $0$ & $0$ & $0$ & $1$ & $\Phi_0$ & ind. & $-1$\\
         $\mathcal B$ & $0$ & $0$ & $1$ & $0$ & $0$ & $0$ & $1$ \\
         $\mathcal C$ & $0$ & $1$ & $0$ & $0$ & $0$ & $0$ & $\frac{1}{2}$ \\
         $\mathcal D$ & $-1$ & $0$ & $0$ & $0$ & $0$ & $0$ & $0$ \\
         $\mathcal E^\pm$ & $0$ & $0$ & $0$ & $0$ & $\pm \sqrt{\frac{3}{k_0}}$ & $0$ & $2$
    \end{tabular}
    \caption{Fixed points for the class 1 model given in Eq.\eqref{eq:model1}. The tag \textit{ind.} states that this parameter is arbitrary, i.e., the row represents a line of fixed points instead of an isolated point.}
    \label{tab:fixed1}
\end{table}

\begin{figure*}
    \centering
    \includegraphics[scale=0.55]{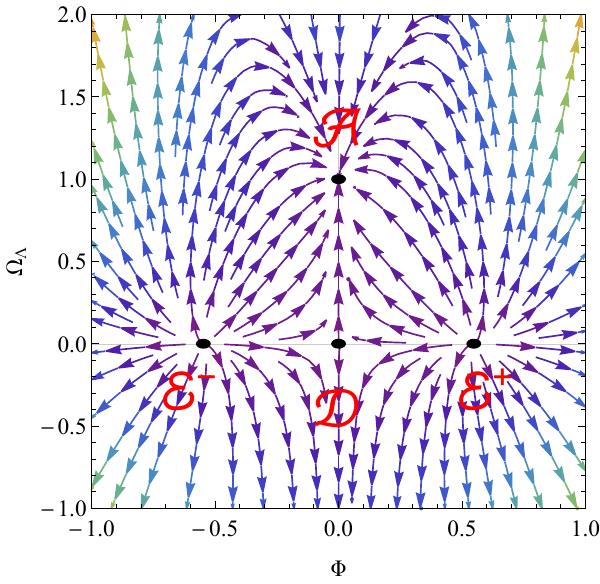}
    \includegraphics[scale=0.55]{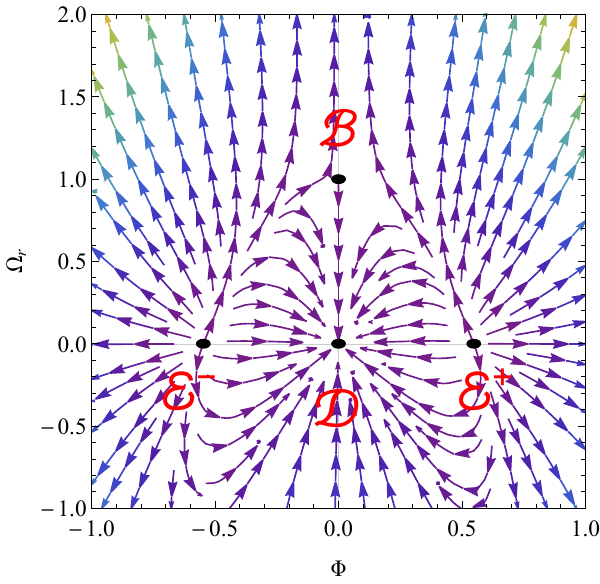}\\
    \includegraphics[scale=0.55]{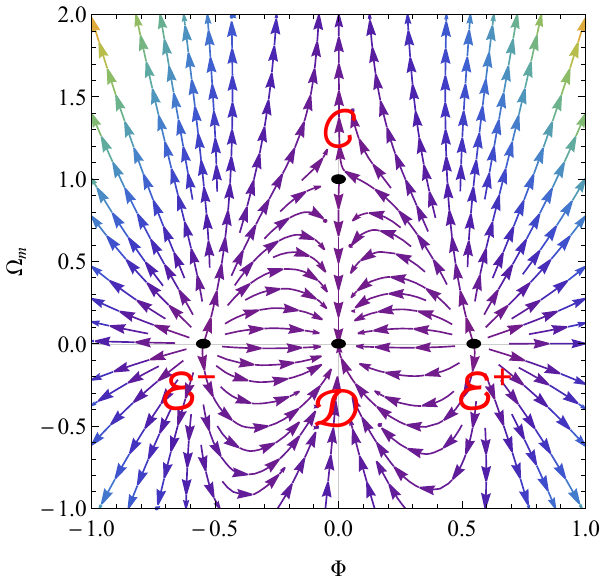}
    \includegraphics[scale=0.55]{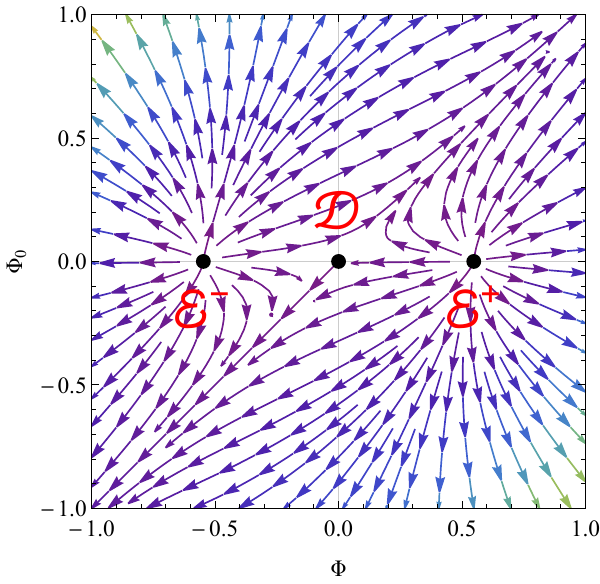}
    \caption{Streamplots of the projected cosmological phase space of the class 1 model in Eq.\eqref{eq:model1} with $k_0=10$ into several submanifolds, namely $\Omega_m=\Omega_r=\Phi_0=0$ (top left), $\Omega_m=\Omega_\Lambda=\Phi_0=0$ (top right), $\Omega_r=\Omega_\Lambda=\Phi_0=0$ (bottom left), and $\Omega_m=\Omega_r=\Omega_\Lambda=0$ (bottom right).}
    \label{fig:streamA}
\end{figure*}

Due to the large dimensionality of the dynamical system, to analyze the stability of the fixed points and the trajectories on the phase space we perform several adequate projections into submanifolds of the phase space for a constant value of $k_0$. Since the projection of the dynamical system into an invariant submanifold results into another dynamical system of a lower dimension, we choose to perform these projections into combinations of invariant submanifolds, namely $\Omega_m=\Omega_r=\Phi_0=0$ ($M_1$), $\Omega_m=\Omega_\Lambda=\Phi_0=0$ ($M_2$), $\Omega_r=\Omega_\Lambda=\Phi_0=0$ ($M_3$), and $\Omega_m=\Omega_r=\Omega_\Lambda=0$ ($M_4$). These projections and the associated two-dimensional streamplots can be found in Fig. \ref{fig:streamA}, whereas a summary of the stability of the fixed points can be found in Table \ref{tab:stability1}.

\begin{table*}
    \centering
    \begin{tabular}{c|c|c|c|c|c}
         & $\mathcal A$ & $\mathcal B$ & $\mathcal C$ & $\mathcal D$ & $\mathcal E^\pm$  \\ \hline
        $M_1$ & $\begin{matrix}\lambda_1=-3\\ \lambda_2=-2\end{matrix}$ (A) & X & X & $\begin{matrix}\lambda_1=-2\\ \lambda_2=2\end{matrix}$ (S) & $\begin{matrix}\lambda_1=4\\ \lambda_2=6\end{matrix}$ (R) \\ \hline 
        $M_2$ & X & $\begin{matrix}\lambda_1=-1\\ \lambda_2=0\end{matrix}$ (A) & X & $\begin{matrix}\lambda_1=-2\\ \lambda_2=0\end{matrix}$ (A) & $\begin{matrix}\lambda_1=4\\ \lambda_2=0\end{matrix}$ (R) \\ \hline 
        $M_3$ & X & X & $\begin{matrix}\lambda_1=-1.5\\ \lambda_2=-0\end{matrix}$ (S) & $\begin{matrix}\lambda_1=-2\\ \lambda_2=0\end{matrix}$ (A) & $\begin{matrix}\lambda_1=4\\ \lambda_2=0\end{matrix}$ (R) \\ \hline 
        $M_4$ & X & X & X & $\begin{matrix}\lambda_1=-2\\ \lambda_2=1\end{matrix}$ (S) & $\begin{matrix}\lambda_1=4\\ \lambda_2=3\end{matrix}$ (R) 
    \end{tabular}
    \caption{Fixed points in the cosmological phase space of the class 1 model in Eq.\eqref{eq:model1} with $k_0=10$ for each of the submanifolds $M_i$, with their respective eigenvalues and stability character, where (A) stands for attractor, (R) stands for reppeler, (S) stands for saddle, and X implies that the fixed point is not visible from that submanifold.}
    \label{tab:stability1}
\end{table*}

The stability analysis indicates that points $\mathcal E^\pm$, when present, which only happens for $k_0>0$, are reppelers in all of the projections considered. However, due to the fact that these points do not stand on the intersection of all invariant submanifolds, they can not be independently global reppelers. This is clear from the streamplots of $M_5$, where the fixed line $\mathcal A$ separates the phase space in two independent regions. Nevertheless, any point on the phase space of the submanifolds considered can always be traced backwards to one of the two points $\mathcal E^\pm$. In these points, one has $\Phi_0=0$. Given that $\Phi_0=Q_0/H$, this can be achieved either via $Q_0=0$ or $H\to\infty$, the latter corresponding to a Big Bang scenario with $a=0$ and/or $\dot a\to\infty$. Since $\Phi=Q/H$ is finite at these fixed points, one either has that $Q_0=0$ and $Q\neq 0$ at the fixed point, i.e., the field $Q$ is initially finite, or that both $H\to \infty$ and $Q\to \infty$, implying that $Q$ diverges at the Big Bang instant. 

The analysis also indicates that two fixed points may behave as attractors, namely $\mathcal A$ and $\mathcal D$, although not for every projection. Nevertheless, in the projection $M_1$ we verify that $\mathcal A$ is an attractor for any point in the phase space with $\Omega_\Lambda>0$, whereas in the projections $M_2$ and $M_3$ the point $\mathcal D$ is a local attractor. These results indicate that a trajectory in the phase space with a positive cosmological constant tends to approach a cosmological constant dominated universe, whereas if the cosmological constant vanishes the trajectories can evolve towards a vacuum universe, if the value of $Q$ is small enough, a matter or radiation dominated universes for a fine-tuned initial value of $Q$, or asymptotically approach a divergent $Q$. Finally, it is worth mentioning that for any vacuum initial condition, i.e., in the projection $M_4$, the universe dynamically evolves towards a situation in which $Q=Q_0$.

\subsection{Class 2 model}
\label{classB}

As a second example, consider the model for the function $F(Q)$ given by
\begin{equation}\label{eq:model2}
F(Q)=k_0Q_0\sqrt{Q_0^2-Q^2},
\end{equation}
where $k_0$ and $Q_0$ are constant free parameters. Similarly to the model studied in the previous section, $k_0$ is a dimensionless parameter but $Q_0$ is not, and thus one need to define one extra dynamical variable to describe the latter,
\begin{equation}
\Phi_0=\frac{Q_0}{H},
\end{equation}
which satisfies the same dynamical equation 
\begin{equation}\label{eq:dyn_extra2}
\Phi'_0=\Phi_0\left(1+q\right),
\end{equation}
contributing thus with an additional invariant submanifold $\Phi_0=0$, that generates $\Phi'_0=0$. The three dynamical functions $F_i$ in this case take the forms
\begin{equation}\label{eq:model2_f1}
F_1=-\frac{1}{3}k_0\Phi_0\sqrt{\Phi_0^2-\Phi^2},
\end{equation}
\begin{equation}\label{eq:model2_f2}
F_2=\frac{1}{3}k_0\Phi_0\frac{\Phi}{\sqrt{\Phi_0^2-\Phi^2}},
\end{equation}
\begin{equation}\label{eq:model2_f3}
F_3=\frac{1}{9}k_0\Phi_0\frac{\Phi_0^2}{\left(\Phi_0^2-\Phi^2\right)^{\frac{3}{2}}}.
\end{equation}
Similarly to the previous case, upon replacing Eqs.\eqref{eq:model2_f1} to \eqref{eq:model2_f3} into Eq.\eqref{eq:dyn_eq_gen1}, one verifies that the submanifold $\Phi=0$ becomes an invariant submanifold, as it implies $\Phi'=0$.

Similarly to the previous model, the system of Eqs.\eqref{eq:dyn_eq_gen1} to \eqref{eq:dyn_eq_gen4}, along with Eq.\eqref{eq:dyn_extra2}, features at most five isolated fixed points and a continuous line of fixed points, summarized in Table \ref{tab:fixed2}. Many of the fixed points are exactly the same as in the class 1 model, namely the radiation and matter dominated universes $\mathcal B$ and $\mathcal C$, respectively, as well as the vacuum solution $\mathcal D$. However, some differences arise in points $\mathcal A$ and $\mathcal E^\pm$. Note that $\mathcal E^\pm$ are contained in $\mathcal A$, as by choosing $\Phi_0=\pm\sqrt{\frac{6}{k_0}}$ in $\mathcal A$ one obtains $\mathcal{E}^\pm$. Nevertheless, we write these points separately to emphasize one fundamental difference. Similarly to the previous case, the line $\mathcal A$ represents a cosmological constant dominated universe with an exponentially accelerated expansion. This line is reached when $Q=0$, and the value of $Q_0$ contributes to the value of the cosmological constant, either via an increase or a decrease depending on weather $k_0$ is negative or positive, respectively. As for points $\mathcal E^\pm$, they also represent exponentially accelerated universes but in this case it is solely the scalar $Q_0$ that drives this expansion, as every other matter component vanishes at this fixed point. Again, the fixed point $\mathcal D$ is the only that represents a global property of the phase space, as it is the only one standing in the intersection of all invariant submanifolds.

 \begin{table}
    \centering
    \begin{tabular}{c|c|c|c|c|c|c|c}
          & $K$ & $\Omega_m$ & $\Omega_r$ & $\Omega_\Lambda$ & $\Phi$ & $\Phi_0$ & $q$  \\ \hline 
        $\mathcal A$ & $0$ & $0$ & $0$ & $1-\frac{1}{3}k_0\Phi_0 |\Phi_0|$ & $0$ & ind. & $-1$\\
        $\mathcal B$ & $0$ & $0$ & $1$ & $0$ & $0$ & $0$ & $1$\\
        $\mathcal C$ & $0$ & $1$ & $0$ & $0$ & $0$ & $0$ & $\frac{1}{2}$\\
        $\mathcal D$ & $-1$ & $0$ & $0$ & $0$ & $0$ & $0$ & $0$ \\
        $\mathcal E^\pm$ & $0$ & $0$ & $0$ & $0$ & $0$ & $\pm\sqrt{\frac{3}{|k_0|}}$ & $-1$
    \end{tabular}
    \caption{Fixed points for the class 2 model given in Eq.\eqref{eq:model2}. The tag \textit{ind.} states that this parameter is arbitrary, i.e., the row represents a line of fixed points instead of an isolated point.}
    \label{tab:fixed2}
\end{table}

\begin{figure*}
    \centering
    \includegraphics[scale=0.63]{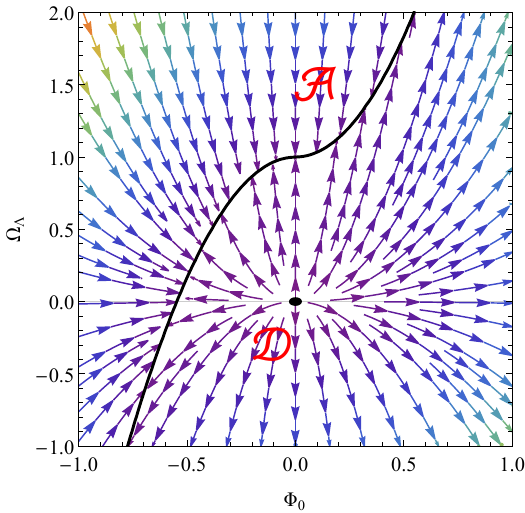}
    \includegraphics[scale=0.63]{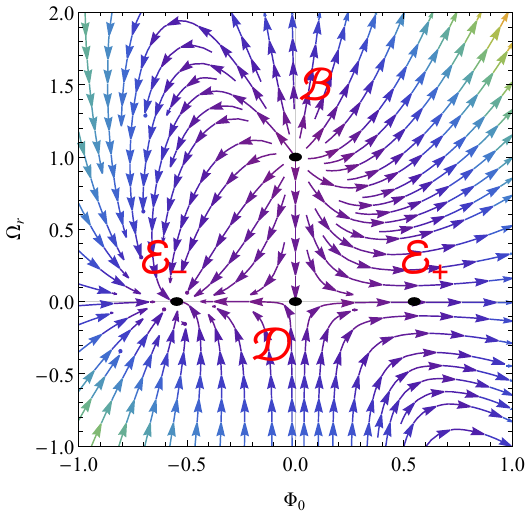}\\
    \includegraphics[scale=0.63]{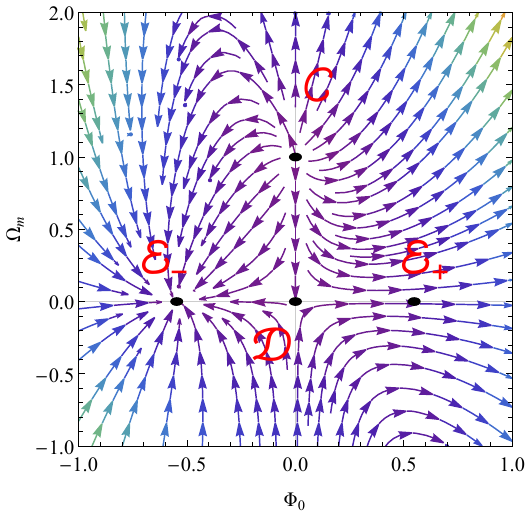}
    \includegraphics[scale=0.63]{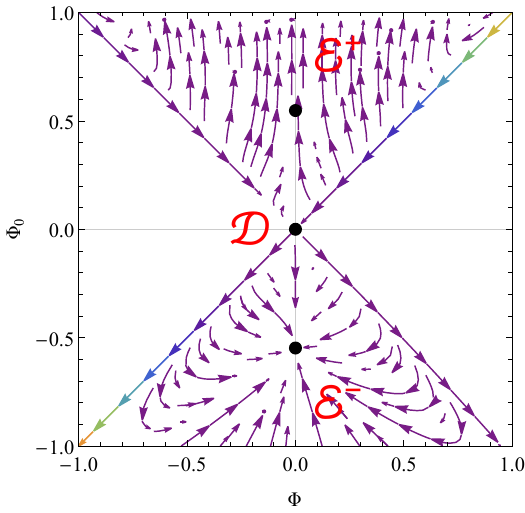}
    \caption{Streamplots of the projected cosmological phase space of the class 2 model in Eq.\eqref{eq:model2} into several submanifolds,  namely $\Omega_m=\Omega_r=\Phi_0=0$ (top left), $\Omega_m=\Omega_\Lambda=\Phi_0=0$ (top right), $\Omega_r=\Omega_\Lambda=\Phi_0=0$ (bottom left), and $\Omega_m=\Omega_r=\Omega_\Lambda=0$ (bottom right).}
    \label{fig:streamB}
\end{figure*}

Similarly to the previous case, due to the large dimensionality of the phase space, we performe several projections into submanifolds of the phase space with a constant $k_0$ to analyze its trajectories and the stability of the fixed points. We again recur to projections into combinations of invariant submanifolds to preserve the character of the dynamical system, namely $\Omega_m=\Omega_r=\Phi_0=0$ ($M_1$), $\Omega_m=\Omega_\Lambda=\Phi_0=0$ ($M_2$), $\Omega_r=\Omega_\Lambda=\Phi_0=0$ ($M_3$), and $\Omega_m=\Omega_r=\Omega_\Lambda=0$ ($M_4$). These projections along with their associated two-dimensional streamplots are given in Fig.\ref{fig:streamB}, whereas a summary of the fixed points and their stability can be found in Table \ref{tab:stability2}. Note that for the $M_4$ projection, only the regions where $|\Phi|\geq|\Phi_0|$ are plotted, as the function $F$ becomes imaginary otherwise. We note that although $\mathcal D$ appears in the projection $M_4$, it does not correspond to a fixed point of that projection since the dynamical system is singular at that point.

\begin{table*}
    \centering
    \begin{tabular}{c|c|c|c|c|c}
         & $\mathcal A$ & $\mathcal B$ & $\mathcal C$ & $\mathcal D$ & $\mathcal E^\pm$  \\ \hline
        $M_1$ & $\begin{matrix}\lambda_1=-2\\ \lambda_2=0\end{matrix}$ (A) & X & X & $\begin{matrix}\lambda_1=2\\ \lambda_2=1\end{matrix}$ (R) & X \\ \hline 
        $M_2$ & X & $\begin{matrix}\lambda_1=2\\ \lambda_2=2\end{matrix}$ (R) & X & $\begin{matrix}\lambda_1=-2\\ \lambda_2=1\end{matrix}$ (S) & $\begin{matrix}\lambda_1=-2\pm 2\\ \lambda_2=1\pm3\end{matrix}$ (A/S) \\ \hline 
        $M_3$ & X & X & $\begin{matrix}\lambda_1=1.5\\ \lambda_2=1\end{matrix}$ (R) & $\begin{matrix}\lambda_1=-1\\ \lambda_2=1\end{matrix}$ (S) & $\begin{matrix}\lambda_1=-1\pm2\\ \lambda_2=1\pm3\end{matrix}$ (A/S) \\ \hline 
        $M_4$ & X & X & X & X & $\begin{matrix}\lambda_1=1\pm3\\ \lambda_2=-2\pm 1\end{matrix}$ (A/S) 
    \end{tabular}
    \caption{Fixed points in the cosmological phase space of the class 2 model in Eq.\eqref{eq:model2} with $k_0=10$  for each of the submanifolds $M_i$, with their respective eigenvalues and stability character, where (A) stands for attractor, (R) stands for reppeler, (S) stands for saddle, and X implies that the fixed point is not visible from that submanifold.}
    \label{tab:stability2}
\end{table*}

The stability analysis reveals, similarly to what has been previously found for the class 1 model, that the trajectories in the phase space tend to evolve towards exponentially expanding universes with $q=-1$. However, whereas in the class 1 model this was only true if one starts from an initial condition with $\Omega_\Lambda>0$, in which case the system evolves to a cosmological constant dominated expansion, in the class 2 model this is true independently of the value of $\Omega_\Lambda$, as long as $\Phi_0$ and $k_0$ have compatible signs. Indeed, in the projections $M_2$, $M_3$, the points $\mathcal B$ and $\mathcal C$ are reppelers, and even though that $\Omega_\Lambda=0$ the system still evolves towards a solution with $q=-1$, namely the points $\mathcal {E}^\pm$ (either one depending on the sign of $k_0$). Also, note that if $\Phi_0=|\Phi|$ with $k_0>0$ or $\Phi_0=-|\Phi|$ with $k_0<0$, the system evolves towards $\Phi=\Phi_0=0$, a singular point in the phase space. Given that $\Phi=Q/H$, this indicates that the system is evolving to a situation where either $\dot a$ diverges or $a=0$, i.e., either a sudden singularity or a Big Crunch. If $\Phi_0=-|\Phi|$ with $k_0>0$ or $\Phi_0=|\Phi|$ with $k_0<0$ instead, the opposite happens: the system evolves asymptotically towards a situation at which $|\Phi|=|\Phi_0|\to\infty$, which indicates that either $\dot a=0$ or $a\to \infty$, i.e., a stationary universe or a Big Chill. Finally, it is interesting to note that if $\Phi_0\neq\pm|\Phi|$, the field $\Phi$ does not evolve in the direction of $\Phi=\Phi_0$ but instead evolves asymptotically towards $\Phi=0$.

An interesting property of the class 2 model is that only a fine-tuned $\Phi_0=0$ initial condition leads to a cosmological constant dominated evolution with $\Omega_\Lambda=1$, as can be seen from the $M_4$ projection. If $\Phi_0\neq 0$, the system evolves towards an exponentially accelerated universe with $\Omega_\Lambda\neq 1$, for which the scalar $\Phi_0$ contributes either positively or negatively, depending on the sign of $k_0$, to the cosmological constant. For example, if $\Phi_0>0$ and $k_0>0$, the system evolves towards a fixed point at which $\Omega_\Lambda<1$, i.e., the field $\Phi$ contributes with a negative cosmological constant that decreases the expansion rate, whereas if $k_0<0$ the system evolves to a fixed point at which $\Omega_\Lambda>1$, i.e., the field $\Phi$ contributes with a positive cosmological constant, thus increasing the expansion rate. This behavior indicates that the scalar field can play the role of a cosmological constant at late times, a situation that we explore in the following sections. 

\subsection{Class 3 model}

As a final example, consider the model for the function $F(Q)$ of the form
\begin{equation}\label{eq:model3}
F(Q)=k_0Q_0^{3/2}\sqrt{Q_0-Q},
\end{equation}
where $k_0$ and $Q_0$ are again constant free parameters. Similarly to the previous models, it is necessary to define one extra dynamical variable to describe the quantity $Q_0$, which takes the usual form
\begin{equation}
\Phi_0=\frac{Q_0}{H},
\end{equation}
which satisfies again the dynamical equation 
\begin{equation}\label{eq:dyn_extra3}
\Phi'_0=\Phi_0\left(1+q\right),
\end{equation}
and corresponding to an additional invariant submanifold $\Phi_0=0$, since in such case $\Phi'_0=0$. The three dynamical functions $F_i$ in this case become
\begin{equation}\label{eq:model3_f1}
F_1=-\frac{1}{3}k_0\Phi_0^{3/2}\sqrt{\Phi_0-\Phi},
\end{equation}
\begin{equation}\label{eq:model3_f2}
F_2=\frac{1}{6}k_0\Phi_0^{3/2}\frac{1}{\sqrt{\Phi_0-\Phi}},
\end{equation}
\begin{equation}\label{eq:model3_f3}
F_3=\frac{1}{36}k_0\Phi_0^{3/2}\frac{1}{\left(\Phi_0-\Phi\right)^{\frac{3}{2}}}.
\end{equation}
In this case, and unlike the two previously studied cases, replacing Eqs.\eqref{eq:model3_f1} to \eqref{eq:model3_f3} into Eq.\eqref{eq:dyn_eq_gen1} does not lead to the appearance of an additional invariant submanifold $\Phi=0$, as in this case this does not imply immediately that $\Phi'=0$.

Unlike the previous two models, the system of Eqs.\eqref{eq:dyn_eq_gen1} to \eqref{eq:dyn_eq_gen4}, along with Eq.\eqref{eq:dyn_extra3} features at most only four isolated fixed points and no continuous line of fixed points, which are summarized in Table \ref{tab:fixed3} . Again, the system features fixes points corresponding to cosmological constant, radiation, and matter dominated solutions, i.e., the points $\mathcal A$, $\mathcal B$, and $\mathcal C$ respectively, with the difference with respect to the previous models being the fact that point $\mathcal A$ here only exists for $\Phi=\Phi_0=0$. This model also features the familiar vacuum solution $\mathcal D$, also present in the previous models. Indeed, all of the fixed points obtained in this model correspond to $\Phi=\Phi=0$, i.e., $Q=Q_0$. Only the point $\mathcal D$ corresponds to a global property of the spacetime as it stands in the intersection of all invariant submanifolds.

 \begin{table}
    \centering
    \begin{tabular}{c|c|c|c|c|c|c|c}
          & $K$ & $\Omega_m$ & $\Omega_r$ & $\Omega_\Lambda$ & $\Phi$ & $\Phi_0$ & $q$  \\ \hline 
        $\mathcal A$ & $0$ & $0$ & $0$ & $1$ & $0$ & $0$ & $-1$\\
        $\mathcal B$ & $0$ & $0$ & $1$ & $0$ & $0$ & $0$ & $1$\\
        $\mathcal C$ & $0$ & $1$ & $0$ & $0$ & $0$ & $0$ & $\frac{1}{2}$\\
        $\mathcal D$ & $-1$ & $0$ & $0$ & $0$ & $0$ & $0$ & $0$
    \end{tabular}
    \caption{Fixed points for the class 3 model given in Eq.\eqref{eq:model3}.}
    \label{tab:fixed3}
\end{table}

\begin{figure*}
    \centering
    \includegraphics[scale=0.63]{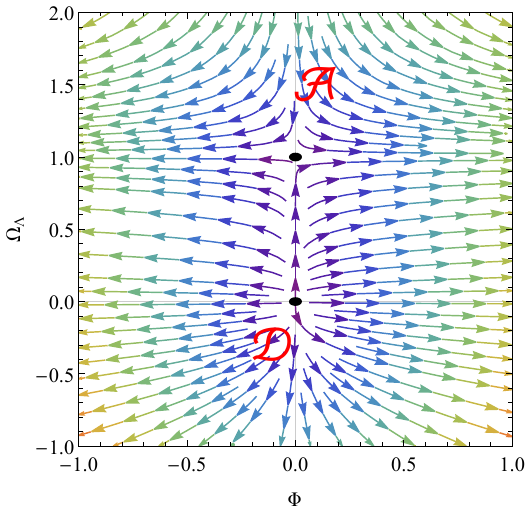}
    \includegraphics[scale=0.63]{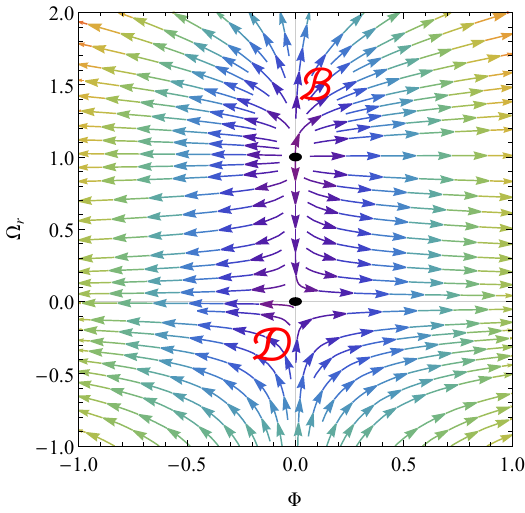}\\
    \includegraphics[scale=0.63]{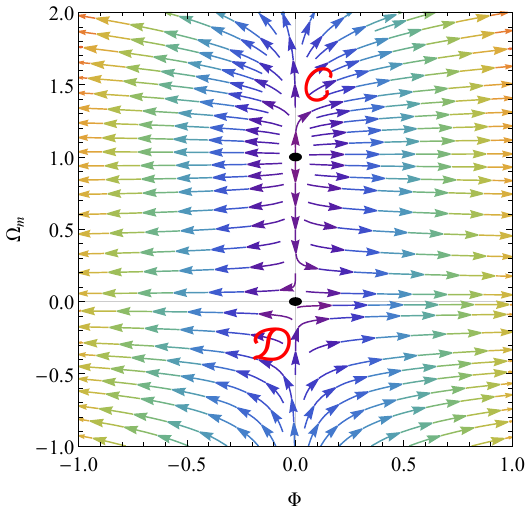}
    \includegraphics[scale=0.63]{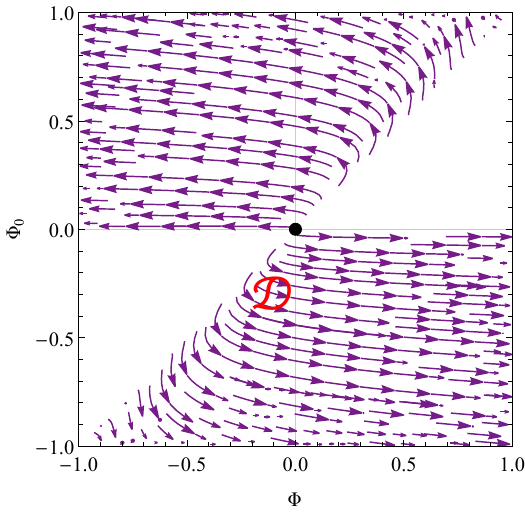}
    \caption{Streamplots of the projected cosmological phase space of the class 3 model in Eq.\eqref{eq:model3} into several submanifolds,  namely $\Omega_m=\Omega_r=\Phi_0=0$ (top left), $\Omega_m=\Omega_\Lambda=\Phi_0=0$ (top right), $\Omega_r=\Omega_\Lambda=\Phi_0=0$ (bottom left), and $\Omega_m=\Omega_r=\Omega_\Lambda=0$ (bottom right).}
    \label{fig:streamC}
\end{figure*}

Similarly to what was done in the previous models, due to the large dimensionality of the phase space, we perform several projections into invariant submanifolds of the phase space with a constant $k_0$ to analyze the trajectories in the phase space and stability of the fixed points. We recur to the same projections as before, namely $\Omega_m=\Omega_r=\Phi_0=0$ ($M_1$), $\Omega_m=\Omega_\Lambda=\Phi_0=0$ ($M_2$), $\Omega_r=\Omega_\Lambda=\Phi_0=0$ ($M_3$), and $\Omega_m=\Omega_r=\Omega_\Lambda=0$ ($M_4$), which preserve the character of the dynamical system. These projections and respective streamplots are given in Fig. \ref{fig:streamC}, and a summary of the fixed points and their stability can be found in Table \ref{tab:stability3}. Similarly to the class 2 model, for the projection $M_4$ we only plot the regions where the functions $F(Q)$ are real. Furthermore, we note that the point $\mathcal D$ does not correspond to a fixed point of that projection because the dynamical system is singular at that point, thus implying that no fixed points are visible in that projection.

\begin{table*}
    \centering
    \begin{tabular}{c|c|c|c|c}
         & $\mathcal A$ & $\mathcal B$ & $\mathcal C$ & $\mathcal D$ \\ \hline
        $M_1$ & $\begin{matrix}\lambda_1=6\\ \lambda_2=-2\end{matrix}$ (S) & X & X & $\begin{matrix}\lambda_1=7\\ \lambda_2=2\end{matrix}$ (R)  \\ \hline 
        $M_2$ & X & $\begin{matrix}\lambda_1=8\\ \lambda_2=2\end{matrix}$ (R) & X & $\begin{matrix}\lambda_1=7\\ \lambda_2=-2\end{matrix}$ (S) \\ \hline 
        $M_3$ & X & X & $\begin{matrix}\lambda_1=7.5\\ \lambda_2=1\end{matrix}$ (R) & $\begin{matrix}\lambda_1=7\\ \lambda_2=-1\end{matrix}$ (S)  \\ \hline 
        $M_4$ & X & X & X & X 
    \end{tabular}
    \caption{Fixed points in the cosmological phase space of the class 3 model in Eq.\eqref{eq:model3} with $k_0=10$  for each of the submanifolds $M_i$, with their respective eigenvalues and stability character, where (A) stands for attractor, (R) stands for reppeler, (S) stands for saddle, and X implies that the fixed point is not visible from that submanifold.}
    \label{tab:stability3}
\end{table*}

Unlike the previous models, the stability analysis reveals that trajectories in the phase space do not evolve towards exponentially accelerated universes in general. Instead, in the more physically relevant regime $0<\Omega_i<1$ for $\Omega_i=\{\Omega_m,\Omega_r,\Omega_\Lambda\}$, all trajectories in the phase space starting from an initial condition with $\Phi\neq 0$ rapidly evolve towards a universe with $\Phi\to\pm\infty$ and $\Omega_i\to 0$. This is visible in the projection $M_1$, where the fixed point $\mathcal A$ is a saddle point, and the point $\mathcal D$ is a reppeler, and also in the projections $M_2$ and $M_3$, where the fixed points $\mathcal B$ and $\mathcal C$ are reppelers and the fixed point $\mathcal D$ is a saddle point. In the particular case $\Phi=0$, trajectories tend to evolve towards a solution with $q=-1$ (point $\mathcal A$ in the projection $M_1$) if $\Omega_\Lambda\neq 0$, or towards a vacuum solution (point $\mathcal D$ in the projections $M_2$ and $M_3$) if $\Omega_\Lambda=0$. This implies that, except for particular fine-tuned situations, the late time cosmic expansion of the universe is not cosmological constant dominated, but instead it approaches some asymptotic state dominated by the scalar field, as we investigate in the sections that follow. This dominance is emphasized by the analysis of the projection $M_4$, where we observe that independently of the values of $\Phi$ and $\Phi_0$ chosen as an initial condition, as long as these correspond to initial states for which the functions $F$ are real, the trajectories in the phase space evolve towards asymptotic configurations with $\Phi\to\pm\infty$ and $\Phi_0\to\pm\infty$, with $\Phi$ diverging quicker than $\Phi_0$.

\section{Cosmological models}
\label{sec:cosmo_models}

In this section, we aim to analyze the hypothesis each class of model considered in the previous section can successfully play the role of dark matter in a cosmological model consistent with the current observations and measurements of the cosmological parameters. For this purpose, we start by rewriting Eq. \eqref{eq:const1} in the more convenient form
\begin{equation}\label{eq:cosmo_const}
1+K=\Omega_m+\Omega_r+\Omega_\Lambda +\Omega_Q,
\end{equation}
where we have defined the scalar field density parameter $\Omega_Q$ as
\begin{equation}
\Omega_Q=F_1-F_2\Phi.
\end{equation}
For each of the classes of models 1 to 3, one can make the following decomposition of $\Omega_{Q}$:
\begin{equation}
\Omega_Q=\Omega_A+\Omega_B,
\end{equation}
where $\Omega_A$ is the collection of terms in $\Omega_Q$ that feature a behavior qualitatively similar to $\Omega_m$, i.e., that contribute to a matter-dominated behavior of the universe, and $\Omega_B$ collectively denotes the remaining terms. Note that the density parameter $\Omega_A$ always preserves the same influence to the cosmological behavior, i.e., an additional matter density, whereas the contribution of $\Omega_B$ varies depending on the explicit form of the function $F(Q)$. 

By way of example, for the class 1 model, integration of the scalar field equation yields $2k_{0}(Q-Q_{0}) = \frac{\xi}{a^{3}}$, for some constant $\xi$, thus yielding 
\begin{align}
\rho_{(\phi)} &=  \frac{1}{8\pi \tilde{G}}\bigg(\xi Q_{0}\frac{1}{a^{3}}+\frac{\xi^{2}}{4k_{0}}\frac{1}{a^{6}}\bigg)
\end{align}
and hence 

\begin{align}
    \Omega_{A} &= \frac{1}{3H_{0}^{2}}\xi Q_{0}\\
    \Omega_{B} &=  \frac{1}{3H_{0}^{2}}\frac{\xi^{2}}{4k_{0}}
\end{align}
A similar decomposition can be made for the classes of models 2 and 3. For each model, if $\Omega_{A}\neq 0$ then $\Omega_{B}\neq 0$. However, the values of $(\Omega_{A},\Omega_{B})$ are independent of one another and hence $\Omega_{B}$ can be made arbitrarily small whilst maintaining and $\Omega_{A}$ of appropriate magnitude as an effective dark matter contribution \footnote{We note that such a bound on the value of $\Omega_B$ does not imply that this parameter must be fine-tuned, since its value can range throughout an infinity of orders of magnitude.}.

To produce cosmological models consistent with the current cosmological measurements from the Planck satellite \cite{Planck:2018vyg} - namely $\Omega_m$, $\Omega_r$, $\Omega_\Lambda$, and $q$ - in what follows we adopt certain specific values for the present density parameters. Given that the universe is observed to be spatially flat, we take $K=0$ which, since it corresponds to an invariant submanifold of the phase space, remains true throughout the entire time evolution. Furthermore, the radiation density parameter is taken to be $\Omega_r=5\times 10^{-5}$. The observed matter density parameter is of the order $\Omega_m\sim 0.3$. However, this value already includes contributions form both baryonic matter and dark matter. Since we want the scalar field $Q$ to play the role of dark matter, we thus take instead $\Omega_m=0.05$, to cover solely the contributions of baryonic matter, while the remaining matter density is taken into $\Omega_A=0.25$. The density parameters $\Omega_\Lambda$ and $\Omega_B$ are free parameters corresponding to a single degree of freedom. Indeed, upon setting a particular value of $\Omega_B$, the value of $\Omega_\Lambda$ is set by Eq.\eqref{eq:cosmo_const}. The present value of the deceleration parameter $q$ can then be extracted from Eq.\eqref{eq:const2}, which must be of the order $q\sim-0.55$ to be consistent with the cosmological observations. 

Upon setting the initial conditions described above, the dynamical system of equations respective to each of the classes of models analyzed in this work can be numerically integrated to obtain a suitable cosmological model. In particular, we are looking for cosmological models featuring a past radiation and matter dominated eras, and currently under a transition into a cosmological constant dominated era.

\subsection{Class 1 model}

\begin{figure*}
    \centering
    \includegraphics{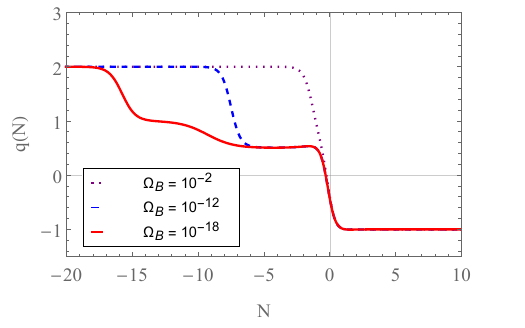}
    \includegraphics{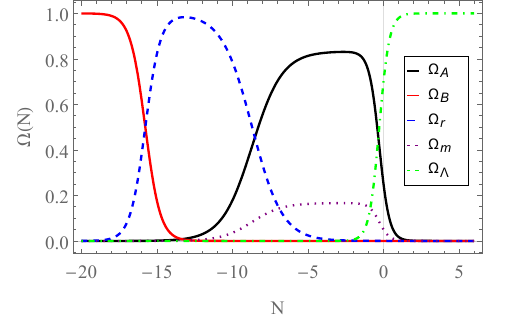}
    \caption{Numerical integration of the deceleration parameter $q(t)$ for the class 1 model given in Eq.\eqref{eq:model1} under initial conditions consistent with the measurements of the Planck satellite, for different values of $\Omega_B(0)$ (left panel) and the energy density parameters $\Omega_i$ for $\Omega_B(0)=10^{-18}$ (right panel). }
    \label{fig:sol_cosmo_AB}
\end{figure*}

Let us first analyze the class 1 model given in Eq.\eqref{eq:model1}. An integration of the dynamical equations under the initial conditions described leads to different cosmological behaviors depending on the choice of present value of $\Omega_B$, see the left panel of Fig.\ref{fig:sol_cosmo_AB}. Indeed, for $\Omega_B\gtrsim 10^{-3}$, one observes that the scalar field dominates throughout most of the past cosmological evolution, which then undergoes a transition towards a cosmological constant dominated epoch in the future. For smaller values $10^{-15}\lesssim \Omega_B\lesssim 10^{-3}$ the scalar field dominance is pushed towards the past and one obtains a cosmological model with an additional matter-dominated phase before the transition towards a cosmological constant dominated epoch. Finally, for even smaller values $\Omega_B\lesssim 10^{-15}$, one obtains a cosmological model with four different expansion phases, starting from a scalar field dominated expansion at early times, followed by radiation and matter dominated eras, and eventually transitioning into a cosmological constant dominated phase. 

For the choice of the present value $\Omega_B=10^{-18}$, the evolution of the density parameters $\Omega_A$, $\Omega_B$, $\Omega_m$, $\Omega_r$, and $\Omega_\Lambda$ is given in the right panel of Fig.\ref{fig:sol_cosmo_AB}, which emphasizes how $\Omega_A$ successfully plays the role of a dark matter component. Indeed, these results indicate that it is always possible to choose the present value of $\Omega_B$ small enough as to allow relativistic MOND theories to reproduce a cosmological behavior consistent with the measurements from the Planck satellite without the necessity of taking into account a dark matter component. Furthermore, the duration of the radiation dominated era can be indefinitely extended backwards to earlier times by choosing smaller values of $\Omega_B$.

\subsection{Class 2 model}

\begin{figure*}
    \centering
    \includegraphics{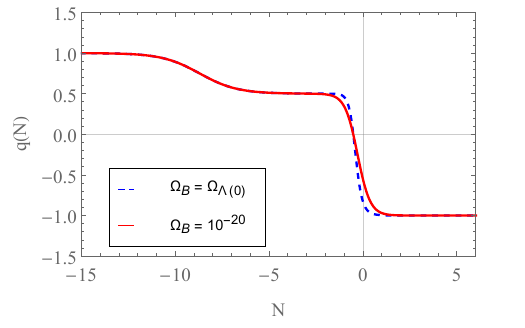}
    \includegraphics{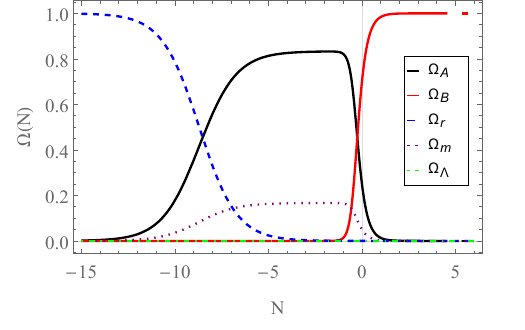}
    \caption{Numerical integration of the deceleration parameter $q(t)$ for the class 2 model given in Eq.\eqref{eq:model2} under initial conditions consistent with the measurements of the Planck satellite, for different values of $\Omega_B(0)$ (left panel) and the energy density parameters $\Omega_i$ for $\Omega_B(0)=\Omega_{\Lambda(0)}$ (right panel).}
    \label{fig:sol_cosmo_AB2}
\end{figure*}

Let us now analyze the class 2 model given in Eq.\eqref{eq:model2}. For this model, one observes that a change in the value of $\Omega_B$ only slightly alters the qualitative behavior of the deceleration parameter over time, see the left panel of Fig.\ref{fig:sol_cosmo_AB2}. In particular, the value of $\Omega_B$ can be given an arbitrarily small value, from which one consequently obtains $\Omega_\Lambda\sim 0.7$, and resulting in a cosmological model perfectly consistent with the $\Lambda\mathrm{CDM}$ model. However, an interesting outcome of this model is the fact that not only $\Omega_A$ can play the role of dark matter, but also $\Omega_B$ can play the role of dark energy. Indeed, taking $\Omega_B=1-\Omega_m-\Omega_r-\Omega_A\equiv \Omega_{\Lambda(0)}$, from which one obtains $\Omega_\Lambda=0$, the cosmological behavior of the deceleration parameter is qualitatively the same, although the transition between the matter and cosmological constant dominated eras becomes more abrupt.

For the choice of the present value $\Omega_B=\Omega_{\Lambda(0)}$, the evolution of the density parameters $\Omega_A$, $\Omega_B$, $\Omega_m$, $\Omega_r$, and $\Omega_\Lambda$ is given in the right panel of Fig.\ref{fig:sol_cosmo_AB2}, from which one observes that $\Omega_\Lambda$ vanishes for the whole time evolution, as expected given the fact that $\Omega_\Lambda=0$ corresponds to an invariant submanifold of the phase space. Instead, it is the density $\Omega_B$ that rises in the late-time cosmic history, leading to a cosmological constant dominated expansion. Also, unlike the class 1 model, the radiation density $\Omega_r$ remains dominant at early times. We note that this model at the background level is identical to a Chaplygin gas unified dark matter-dark energy model and it is likely that if it is to additionally play the role of dark energy (rather than its equation of state being extremely close to zero for the span of the universe's evolution), the model will encounter problems in accounting for the power spectrum of matter density perturbations \cite{Sandvik:2002jz}. We emphasize that if one restricts the analysis to reproduce solely the behavior of CDM and not dark matter, the issue mentioned above does not arise. 

\subsection{Class 3 model}

\begin{figure*}
    \centering
    \includegraphics{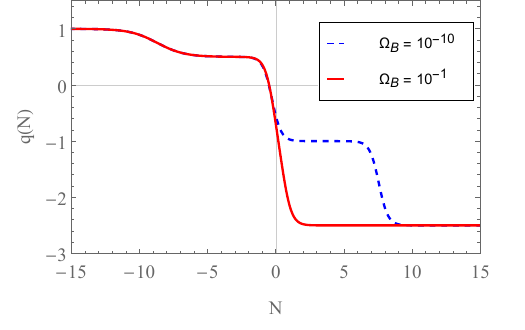}
    \includegraphics{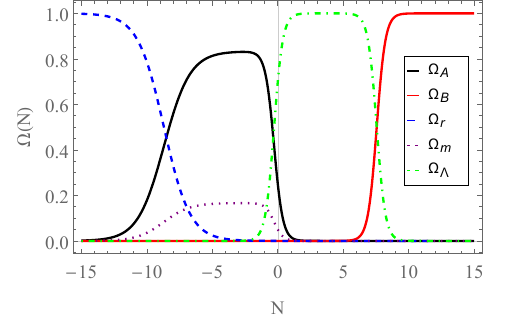}
    \caption{Numerical integration of the deceleration parameter $q(t)$ for the class 3 model given in Eq.\eqref{eq:model3} under initial conditions consistent with the measurements of the Planck satellite, for different values of $\Omega_B(0)$ (left panel) and the energy density parameters $\Omega_i$ for $\Omega_B(0)=10^{-10}$ (right panel).}
    \label{fig:sol_cosmo_AB3}
\end{figure*}

Finally, let us analyze the class 3 model given in Eq.\eqref{eq:model3}. For this model, again different qualitative cosmological behaviors can arise depending on the value of $\Omega_B$, see the left panel of Fig.\ref{fig:sol_cosmo_AB3}. Indeed, for larger values $\Omega_B>10^{-2}$, one observes a direct transition between the matter dominated era into an accelerated scalar field dominated era with $q=-5/2$, whereas for smaller values $\Omega_B<10^{-2}$ a finite period of cosmological constant domination arises before the transition into an accelerated scalar-field domination. 

Taking the present value $\Omega_B=10^{-10}$, the evolution of the density parameters $\Omega_A$, $\Omega_B$, $\Omega_m$, $\Omega_r$, and $\Omega_\Lambda$ is given in the right panel of Fig.\ref{fig:sol_cosmo_AB3}. Similarly to the class 2 model, one observes that the radiation density $\Omega_r$ is always dominant at early times and that $\Omega_A$ successfully plays the role of the dark matter component. The main difference with respect to the previous two models is that the cosmological constant dominated phase is ephemeral, and at late times it is replaced by a domination of $\Omega_B$. Nevertheless, note that the present value of $\Omega_B$ can be made arbitrarily small as to postpone the future scalar-field dominated phase to arbitrarily later times, while preserving the past cosmological behavior consistent with the current observations.

Finally, it is interesting to note that the future asymptotic behavior of the cosmological solutions in this model are described by a value of $q=-2.5$, thus corresponding to a case of phantom matter with $w<-1$. A property of cosmological models populated by this type of matter is that the scale factor diverges in a finite time interval, i.e., $a\left(t\to t_s\right)\to \infty$, where $t_s$ is the instant at which the divergence occurs, also known as a Big Rip scenario. Indeed, the evolution of the scale factor with time for a constant value of $q=q_0$ can be written in the form
\begin{equation}
    a\left(t\right)=a_0\left[1+H_0\left(1+q_0\right) t\right]^{\frac{1}{1+q_0}},
\end{equation}
where the limit $q\to -1$ corresponds to an exponential expansion $a\left(t\right)=a_0\exp\left(H_0 t\right)$ by virtue of $\exp\left(x\right)=\lim_{n\to\infty}\left(1+\frac{x}{n}\right)^n$. Thus, one verifies that if $q_0<-1$, or $q_0+1<0$, the exponent $\frac{1}{1+q_0}$ becomes negative. Furthermore, since $H_0$ has been observed to be positive, this implies that $a\left(t\right)$ diverges at the instant $t_s=-\frac{1}{H_0\left(1+q_0\right)}$. Inserting the current observed value of $H_0$ and $q_0=-2.5$ into the expression above for $t_s$, one obtains a result of $t_s\sim 3.04 \times 10^{17} \text{s}$ for the time at which the Big Rip occurs. Once the transition between the cosmological constant dominated and the scalar field dominated eras is completed, the obtained value of $t_s$ thus corresponds to an upper bound on the time necessary to reach a Big Rip scenario. We refer the reader to Ref. \cite{Caldwell:2003vq} for more details regarding the Big Rip scenario. 

\section{Generalizations and implications for the weak field limit}
\label{sec:generalizations}
It is clear that for each of the previous models, parameters and initial data can be chosen such that the expansion history of the universe closely resembles to that of the $\Lambda\mathrm{CDM}$ model, where the deviation from dark matter behaviour of $\rho_{(\phi)}$, parameterized by the parameter $\Omega_{B}$, can be made arbitrarily small. We will now show, however, that smaller values of $\Omega_{B}$ (and hence smaller values of $w_{(\phi)}$ for a longer span of cosmic time) can be in tension with astrophysical constraints on the AeST theory.

Consider the following generalizations of the functions considered up to this point, which in the following we refer to as $\alpha$ and $\beta$ models: 
\begin{align}
    F_{(\alpha,n)}(Q) &= \frac{\alpha_{n}}{2}Q_{0}^{2(1-n)}(Q^{n}-Q^{n}_{0})^{2} ,\label{Falphan}
\end{align}
\begin{align}
F_{(\beta,n)}(Q) &= -\frac{\beta_{n}}{2}Q_{0}^{2-\frac{n}{2}}\sqrt{Q_{0}^{n}-Q^{n}} \label{Fbetan},
\end{align}
where $n$ is assumed to be a positive integer. For example, for the $\alpha$ models when $n=1$ we recover the class 1 model and for $n=2$ we recover a model which is identical in FLRW symmetry to Scherrer K-essence model \cite{Scherrer2004} (which is also a limiting form of the Ghost Condensate scalar field model \cite{ArkaniHamedEtAl2003}), For the $\beta$ models when $n=1$ we recover the class 3 model and for $n=2$ we recover the class 2 model.

For the $\alpha$ models we can define a new field $V=(Q^{n}-Q_{0}^{n})/Q_{0}^{n}$ and constants $\xi_{n}=\frac{1}{3}\alpha_{n}Q_{0}^{2}/H_{0}^{2}$ and $\zeta_{n}$, such that the Friedmann equation and integrated scalar field equation of motion become: 
\begin{align}
\frac{H^{2}}{H_{0}^{2}} &= \Omega_{\Lambda}+\frac{\Omega_{b}}{a^{3}}+\frac{\Omega_{r}}{a^{4}}+\xi_{n}\bigg(nV +\frac{(2 n-1)}{2} V^2\bigg) \label{alpha_fried},\\
\frac{\zeta_{n}}{a^{3}} &= \xi_{n}V \left(1+V\right)^{\frac{n-1}{n}} .\label{alpha_eqmo}
\end{align}
Similarly, for the $\beta$ models we can define a new field $W = \frac{Q_{0}^{n}-Q^{n}}{Q_{0}^{n}}$, and constants $\chi_{n} =  \frac{1}{12}\beta_{n}\frac{Q_{0}^{2}}{H_{0}^{2}}$ and  $\nu_{n}$ such that the Friedmann equation and integrated scalar field equation of motion become:
\begin{align}
\frac{H^{2}}{H_{0}^{2}} &= \Omega_{\Lambda}+\frac{\Omega_{b}}{a^{3}}+\frac{\Omega_{r}}{a^{4}}+\chi_{n}\bigg(\frac{n  }{ \sqrt{W}}+2\bigg(1-\frac{n}{2}\bigg)
   \sqrt{W} \bigg)\label{beta_fried},\\
\frac{\nu_{n}}{a^3} &=\frac{\chi_{n}(1-W)^{\frac{n-1}{n}}}{ \sqrt{W}}. \label{beta_eqmo}
\end{align}
These equations can be used to determine the background cosmic evolution, which is described by $(a(t),V(t))$ and $(a(t),W(t))$ respectively. Following the arguments of Section \ref{sec:theory}, the effective density of the field $\phi$ approximates that of dark matter if, respectively, $|V|\ll 1$ and $|W|\ll 1$, and the freedom of choice of parameters $(\xi_{n},\zeta_{n})$ and $(\chi_{n},\nu_{n})$ can be used to yield an appropriate abundance of the effective dark matter and an arbitrarily small correction to dust-like behaviour over the range of the scale factor $a(t)$ of interest. Indeed, the scalar field equation of state and adiabatic sound speed for $\alpha$ and $\beta$ models are given by:
\begin{align}
w_{(\alpha)}&= \frac{V}{2n+(2n-1)V}, \label{alpha_w}\\
 c_{ad(\alpha)}^{2} &= \frac{V}{n+(2n-1)V} ,\label{alpha_csq}\\
w_{(\beta)} &= -\frac{2 W}{n+(2-n) W} ,\label{beta_w}\\
c_{ad(\beta)}^{2} &=\frac{2 W}{n + (n-2) W}.\label{beta_csq} 
\end{align}
In Table \ref{tab:asymptotic_w_table}, the general asymptotic scaling of energy density at early and late cosmic times (defined respectively as $a\rightarrow 0$ and $a\rightarrow \infty$, however parameters may take values so that scalar field evolution at the present day $a=a_{0}$ already strongly approximates one of these limiting forms) for the models is shown for cosmological solutions containing a span of scale factors where the gravitational effect of the field $\phi$ approximates that of dust.

\begin{widetext}
\begin{table*}[htbp]
    \centering
    \begin{tabular}{|c|c|c|}
        \hline
        $F(Q)=-\frac{1}{2}{\cal F}(0,Q)$ & Early-time scaling of energy density & Late-time scaling of energy density \\
        \hline
               $(Q^{n}-Q^{n}_{0})^{2}$   & $a^{-\frac{6n}{2n-1}}$& $a^{-3}$ \\
                    $\sqrt{Q_{0}^{n}-Q^{n}}$ & $a^{-3}$ & $n=1: a^{3}$ , \quad $n>1:\mathrm{cst.}$\\
        \hline
    \end{tabular}
    \caption{Table illustrating various functional forms of $F(Q)$ alongside accompanied asymptotic scaling of their contribution to the cosmic energy density at early and late cosmic times.}
    \label{tab:asymptotic_w_table}
\end{table*}
\end{widetext}

The freedom in choice of parameters and initial data of fields allows $(w_{(\alpha)},w_{(\beta)})$ to be arbitrarily small for relevant values of $a$. This additionally applies to the adiabatic sound speeds $(c_{ad(\alpha)}^{2},c_{ad(\beta)}^{2})$. This is an important result as the behaviour of cosmological perturbations to the field $\phi$ has much in common with with so-called Generalized Dark Matter (GDM) models i.e. fluid models of dark matter with a non-vanishing equation of state and squared adiabatic sound speed of perturbations \cite{SkordisZlosnik2020}. Hence, smallness of $(w_{(\phi)},c_{ad(\phi)}^{2})$ makes it likely that not only the cosmic background but also the influence of the new AeST degrees of freedom on perturbations to the cosmic microwave background can resemble that of CDM.

\begin{figure}[h!]
    \includegraphics[height=5cm]{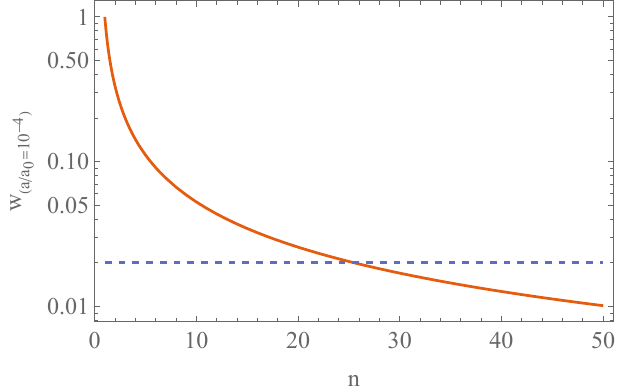}
    \caption{Plot of the equation of state $w_{(a/a_{0}=10^{-4})}$ as a function of $n$ for $\alpha$ models subject to the constraint that $\mu^{2}=\mu_{*}^{2}$. The dotted line denotes the constraint on the maximum value of the GDM equation of state at $a/a_{0}=10^{-4}$ found in \cite{KoppEtAl2018}.}
    \label{fig:wofn}
\end{figure}

However, we will now show that for many models, the span of cosmic time for which the $\phi$ field approximates pressureless dust can be in tension with the recovery of modified gravity-like phenomenology in astrophysical systems. The action of the AeST theory in the quasistatic weak field limit is given by:
\begin{align}
S[\Phi,\varphi] &= - \int d^{4}x \bigg(\frac{2-K_{B}}{16\pi \tilde{G}}\bigg[|\vec{\nabla}\Phi|^{2}-2\vec{\nabla}\Phi\cdot \vec{\nabla}\varphi + |\vec{\nabla}\varphi|^{2}\nonumber\\
&-\mu^{2}\Phi^{2}+{\cal J}\bigg(\vec{\nabla}\varphi\cdot \vec{\nabla}\varphi\bigg)\bigg]+\Phi\rho\bigg),
\end{align}
where $\vec{\nabla}$ is the spatial gradient operator, $\Phi$ is the Newtonian gravitational potential, and $\varphi$ is the deviation of the scalar field from the contemporary asymptotic cosmological value $\phi \sim Q_{0}t$, and $\rho$ is the density of baryonic matter. In the absence of the term proportional to $\mu^{2}$, a multi-field generalization of the modified Poisson equation in Eq. (\ref{modified_poisson}) is recovered. The effect of the term $-\mu^{2}\Phi^{2}$ is to provide a maximum length scale $\sim \mu^{-1}$ over which MOND phenomenology can apply before a different regime -  where the `mass term' $\mu^{2}\Phi^{2}$ dominates - is encountered \cite{Verwayen:2023sds}. This is an important constraint as there appears to be evidence for MOND phenomenology up to scales of around $1 \mathrm{Mpc}$ \cite{Banik:2021woo}. For $\alpha$ and $\beta$ models we have that the quantity $\mu^{2}$ is given by:
\begin{align}
\mu^{2}_{(\alpha)} &=\frac{3 H_{0}^2 \xi_{n} }{2 (2-K_{B})}\left(
2n^{2}+6n^{2}V_{(a=a_{0})}+(4n^{2}-1)V_{(a=a_{0})}^{2}\right),\\
 \mu^{2}_{(\beta)} &= \frac{3 H_{0}^{2} \chi_{n}}{2 (2-K_{B}) }\frac{\left(n^{2}-(n^{2}-4)W_{(a=a_{0})}^{2}\right)}{W_{(a=a_{0})}^{3/2}}.
\end{align}
The value of $H_{0}^{-1}$ is around $(3/h) \mathrm{Gpc}$ 
(where $h$ is the dimensionless Hubble parameter, which is of order unity) so the successful recovery of MOND phenomenology implies that $\mu^{2}$ should have a maximum value of around $(9/h^{2})10^{6}H_{0}^{2} \equiv \mu_{*}^{2}$.
Consider, for example, an $\alpha$ model parameterized by a number $n$. Assuming that the parameter $K_{B}\ll 1$, letting $\mu_{\alpha}^{2}=\mu^{2}_{*}$ imposes a constraint on the parameters $(\xi_{n},V(a=a_{0}))$. Furthermore, requiring that the dust-like contribution to the critical density of matter in Eq. (\ref{alpha_fried}) is equal to the inferred dark matter density $\Omega_{dm}$ fixes both of the $(\xi_{n},V(a=a_{0}))$ and hence determines $\zeta_{n}$ from the scalar field equation of motion Eq. (\ref{alpha_eqmo}). This in turn enables determination of $V(a)$ for general $a$ and hence $(w_{(\phi)}(a),c_{ad(\phi)}^{2}(a))$ from Eqs. (\ref{alpha_w}) and (\ref{alpha_csq}).
In \cite{KoppEtAl2018}, GDM models with a scale factor-dependent parameterization of $(w(a),c_{ad}^{2}(a))$ were compared to a wide variety of data from precision cosmology, with the resulting constraint that for these GDM models, the equation of state at scale factor $a/a_{0}=10^{-4}$ should be of the order $w\leq 2\times 10^{-2}$. We adopt that as a constraint on $w_{(\phi)}$ at this value of scale factor. A plot of $\alpha$ models with $\mu^{2}=\mu^{2}_{*}$ is given in Figure \ref{fig:wofn} where it can be seen that $n$ must be larger than of the order $25$ to be consistent with constraints on $\mu^{2}$ and $w_{(a/a_{0}=10^{-4})}$.  
A model such as $F(Q) = \alpha_{e} (e^{(Q-Q_{0})^{2}/Q^{2}_{0}}-1)$  is also consistent with data for the same reason that models with large $n$ are \cite{SkordisZlosnik2020}. Therefore the class 1 model ($n=1$) is likely inconsistent with the requirement that the model successfully account for dark matter effects in cosmology and - via modified gravity effects - on smaller, astrophysical scales.

A similar process can be used to derive quantities determining cosmic evolution in the 
$\beta$ models subject to the constraint $\mu^{2}_{(\beta)}=\mu^{2}_{*}$ and that the models account for all of the effective cosmological dark matter. It is found that for class 3 ($n=1$) and class 2 ($n=2$) models that $c_{ad(a=a_{0})}^{2} = {\cal O}(10^{-8})$ and $w_{(a=a_{0})} = {\cal O}(-10^{-8})$ with these values increasing with higher, positive values of $n$. Given that $w$ and $c_{ad}^{2}$ \emph{decrease} in magnitude for smaller values of $a$, then the GDM constraints are automatically satisfied. The smallness of $(w(a),c_{ad}^{2}(a))$ 
suggests that the models of class 2 and 3 may be viable as alternatives to dark matter in cosmology at the background and perturbative level while satisfying astrophysical constraints produced by the limiting size of $\mu$.

\section{Conclusions}\label{sec:concl}

In this work, we have used the dynamical system approach to analyze the cosmological phase space of relativistic AeST theories of gravity and their suitability as alternatives to the CDM scenario. For this purpose, we have studied the phase space structure in terms of critical points and invariant submanifolds for three different models (class 1, class 2, and class 3), and we have performed full numerical integrations of the system of dynamical equations subjected to initial conditions consistent with the cosmological parameters inferred from data from the Planck satellite.

For the class 1 model, we verified that several well-known cosmological behaviors, namely matter, radiation, cosmological constant, and scalar field dominated expansions, correspond to equilibrium states of the phase space. Furthermore, while matter and radiation dominated scenarios are meta-stable (saddle points in the phase space), cosmological constant dominated scenarios are stable (attractors), whereas scalar field dominated ones are unstable (reppelers). A numerical integration of the dynamical equations shows that cosmological models consistent with the measurements from the Planck satellite and with a positive cosmological constant density $\Omega_\Lambda$ always evolve asymptotically towards an exponentially accelerated solution, thus confirming the stable character of such a configuration. On the other hand, an integration backwards in time shows that the past behavior of the cosmological solution strongly depends on the value of $\Omega_B$ - a number which parameterizes the deviation of the scalar field energy density from dust-like behaviour. Indeed, for large values of $\Omega_B$, one observes that a backwards integration reveals a direct transition from a scalar field dominated era into a cosmological constant dominated era, thus skipping through the expected radiation and matter dominated eras. These behaviors can however be recovered by decreasing the value of $\Omega_B$, which can push the scalar-field dominated era backwards in time indefinitely, while keeping the present cosmological parameters consistent with the observations.

The fixed point structure for the class 2 model features the same number of fixed points as that of the class 1 model, representing the same well known cosmological behaviors, but with important differences in the scalar field contribution. Indeed, in this model the scalar field dominated era is a future feature of the cosmological evolution, instead of a past one. Furthermore, the scale factor during the scalar field dominated era presents the same behavior as the cosmological constant dominated era. As a result, the cosmological constant dominated fixed point changes its stability character from attractor to saddle point, and it does not require $\Omega_\Lambda=1$ due to the scalar field contribution. Given that the scalar field contribution appears mostly in the future, the entire past cosmological history features the required radiation and matter dominated eras, before transitioning into a cosmological constant or scalar field dominated eras. For this class of model, the value of $\Omega_B$ controls how abrupt this transition is, with small values of $\Omega_B$ maintaining the present deceleration parameter consistent with the current measurements and larger values of $\Omega_B$ pushing this parameter to larger negative values. 

Finally, for the class 3 model, none of the fixed points corresponds to a scalar field dominated era. This happens not because the scalar field is never dominant, but because this domination is achieved asymptotically in the future, at the boundary of the phase space $\Phi\to \pm\infty$. Consequently, a backwards integration of the cosmological equations shows both a radiation and a matter dominated eras, followed by a transition into either a temporary cosmological constant dominated era or directly into a scalar field dominated era. The parameter $\Omega_B$ again controls this transition and the duration of the cosmological constant dominated era. Indeed, for large values of $\Omega_B$ the cosmological constant dominated era is nonexistent and a transition from a matter dominated to a scalar field dominated era occurs directly, while for small values of $\Omega_B$ the transition from a cosmological constant dominated and a scalar field dominated era can be pushed indefinitely into the future, thus preserving the present cosmological parameters consistent with the experimental observations. It is also interesting to note that the future asymptotic state of this model corresponds to a case of phantom matter, eventually leading the cosmological solution into a Big Rip scenario.

Finally we considered a class of generalizations of the previous models. It was shown that a large variety of these models (including the class 1 model) are incompatible with the requirement that the effective equation of state of the scalar field is small enough at early cosmic times whilst simultaneously allowing for MOND phenomenology across a sufficiently wide of span of astrophysical length scales. Preliminary results indicate that class 2 and class 3 models could be consistent with these constraints whilst additionally acting as a realistic alternative to dark matter at the level of cosmological perturbations.

Our results thus show that a wide class of functions $F(Q)$ can successfully reproduce meaningful and physically relevant cosmological background evolution, even in the absence of a CDM component. Nevertheless, our results also imply that relativistic MOND theories in the class 1 and 3 models are unable to play the role of a cosmological constant, as the cosmological phase space features an invariant submanifold at $\Omega_\Lambda=0$, effectively preventing any cosmological solution to evolve towards an exponentially accelerated expansion driven solely by the scalar field. Furthermore, and even though the phase space of the class 2 model features stable critical points where the expansion is exponentially accelerated even in the absence of a cosmological constant density, it is likely that none are consistent with data. Nevertheless, we emphasize that if one focus solely in the modelling of CDM and not dark energy, all three models can successfully reproduce the observational constraints while playing the role of CDM. 

\begin{acknowledgments}
JLR acknowledges the European Regional Development Fund and the programme Mobilitas Pluss for
financial support through Project No.~MOBJD647. This work is part of the project No.~2021/43/P/ST2/02141 co-funded by the Polish National Science Centre and the European Union Framework Programme for Research and Innovation
Horizon 2020 under the Marie Sk\l{}odowska-Curie grant agreement No. 945339.
\end{acknowledgments}

\bibliography{main}
\end{document}